\documentclass[default,iicol]{sn-jnl}

\usepackage{bbding}

\jyear{2023}%

\theoremstyle{thmstyleone}%
\theoremstyle{thmstyletwo}%

\theoremstyle{thmstylethree}%

\begin{document}
\setlength{\heavyrulewidth}{1.5pt}
\setlength{\abovetopsep}{4pt}

\title[The pyPPG Toolbox]{pyPPG: A Python toolbox for comprehensive photoplethysmography signal analysis}


\author[1]{\fnm{Márton Á.} \sur{Goda}} 

\author[2]{\fnm{Peter H.} \sur{Charlton}}

\author[1]{\fnm{Joachim A.} \sur{Behar}}

\affil[1]{\orgdiv{Faculty of Biomedical Engineering}, \orgname{Technion Institute of Technology}, \orgaddress{\street{Technion-IIT}, \city{Haifa}, \postcode{32000}, \country{Israel}}}

\affil[2]{\orgdiv{Department of Public Health and Primary Care}, \orgname{University of Cambridge}, \city{Cambridge}, \postcode{CB1 8RN}, \country{United Kingdom}}

\abstract{


Photoplethysmography is a non-invasive optical technique that measures changes in blood volume within tissues. It is commonly and increasingly used for in a variety of research and clinical application to assess vascular dynamics and physiological parameters. Yet, contrary to heart rate variability measures, a field which has seen the development of stable standards and advanced toolboxes and software, no such standards and open tools exist for continuous photoplethysmogram (PPG) analysis. Consequently, the primary objective of this research was to identify, standardize, implement and validate key digital PPG biomarkers. This work describes the creation of a standard Python toolbox, denoted \textit{pyPPG}, for long-term continuous PPG time series analysis recorded using a standard finger-based transmission pulse oximeter. The improved PPG peak detector had an F1-score of 88.19\% for the state-of-the-art benchmark when evaluated on 2,054 adult polysomnography recordings totaling over 91 million reference beats. This algorithm outperformed the open-source original Matlab implementation by $\sim$5\% when benchmarked on a subset of 100 randomly selected MESA recordings. More than 3,000 fiducial points were manually annotated by two annotators in order to validate the fiducial points detector. The detector consistently demonstrated high performance, with a mean absolute error of less than 10 ms for all fiducial points. Based on these fiducial points, \textit{pyPPG} engineers a set of 74 PPG biomarkers. Studying the PPG time series variability using \textit{pyPPG} can enhance our understanding of the manifestations and etiology of diseases. This toolbox can also be used for biomarker engineering in training data-driven models. \textit{pyPPG} is available on physiozoo.org (upon publication.)}.




\maketitle
\onecolumn


\section{Introduction}\label{sec_intro}

Photoplethysmography is an optical sensing technique widely used for health and fitness monitoring in clinical and consumer devices \cite{Charlton2020rev}, such as smartwatches and pulse oximeters. Photoplethysmography was developed in the 1930s \cite{Allen_2007}, and its potential value for assessing cardiovascular health was recognised in the 1940s \cite{Dillon1941}. It was not until the 1970s that photoplethysmography became widely used as the sensing technology in pulse oximeters \cite{Aoyagi2003}. Photoplethysmography-based wearable devices entered the consumer market in the 2010s \cite{Charlton2021d}, and are now used by millions of people for unobtrusive health monitoring \cite{Natarajan2020}.

The photoplethysmogram (PPG) signal contains a wealth of information on the heart, blood vessels, breathing, and autonomic nervous system \cite{Allen_2007}. Consequently, much research is focused on extracting physiological information from the PPG \cite{Mejia_2022}, including physiological parameters such as blood pressure and breathing rate \cite{mukkamala_cuffless_2022,Charlton2017c}, and disease indicators, such as vascular age and cardiovascular risk markers \cite{Charlton2020rev,Charlton_2022assessing}. The value of photoplethysmography is rapidly increasing: its value for heart rate and oxygen saturation monitoring is well established, its utility for detecting atrial fibrillation has recently been demonstrated \cite{Perez2019}, and its potential to detect other diseases such as sleep apneas and peripheral arterial disease is being researched \cite{Charlton2020rev,Charlton_2022assessing}. Despite the widespread and increasing use of photoplethysmography, there is a lack of open-source tools for detailed analysis of the PPG. This paper presents $pyPPG$, an open-source, validated Python toolbox for PPG signal analysis.


\subsection{The photoplethysmogram (PPG) signal}

The PPG signal is an optical measurement of the arterial pulse wave \cite{Charlton2019}, \textit{i.e.}, the wave generated when blood is ejected from the heart, temporarily increasing arterial pressure and causing vessel expansion and contraction \cite{alastruey_arterial_2023}. Consequently, the PPG signal is influenced by a range of physiological systems, such as: the heart, including heart rate, heart rhythm, and the nature of ejection \cite{Charlton2020rev}; the blood vessels, including vessel stiffness, diameter, and blood pressure \cite{Charlton2020rev}; the microvasculature, including peripheral compliance and resistance \cite{Charlton2020rev}; the autonomic nervous system which influences heart rate variability \cite{Gil2010b}; and the respiratory system, which impacts the pulse wave through changes in intrathoracic pressure \cite{charlton_2017extraction}. Thus, there is potential to extract much physiological information from the PPG signal.

The PPG signal can be acquired using a range of sensors and devices. PPG sensors consist of a light source such as an LED, and a light sensor, such as a photodiode \cite{Sun2016}. The light source illuminates a region with vasculature, such as the fingertip, and the light sensor measures how much light is either transmitted through or reflected from the tissue. The amount of light received fluctuates with each heartbeat: usually, the amount of absorbed light increases during systole when blood volume is greatest, and then decreases during diastole when blood volume returns to its initial level \cite{Allen_2007}. This produces a pulse wave bearing several features, which can be interpreted as physiological biomarkers (see Figure \ref{fig:fiducial_points}) \cite{Charlton2020rev}. In pulse oximeters, the PPG is typically acquired in transmission mode using a fingerclip probe \cite{Nitzan2020}, while in consumer devices such as smartwatches, fitness trackers, and earbuds (\textit{i.e. hearables}), it is typically acquired reflection mode \cite{Charlton2021d}.


\subsection{Applications of photoplethysmography}

At present, the most common applications of photoplethysmography are for heart rate monitoring in smartwatches \cite{Temko_2017}, and for oxygen saturation monitoring in pulse oximeters \cite{Nitzan2020}. Smartwatches, fitness trackers and hearables are widely used, with an estimated 1.1 billion connected wearable devices worldwide in 2022 \cite{statista_global_2023}. Pulse oximetry is a standard-of-care technique used in a range of clinical settings from intensive care to home monitoring \cite{Ortega2011, Greenhalgh2021}. Recently, the applications of photoplethysmography-based wearables have been expanded to include atrial fibrillation detection \cite{Perez2019}, blood pressure monitoring \cite{Vybornova2021}, and oxygen saturation monitoring \cite{spaccarotella_assessment_2022}. Several additional potential applications of wearable photoplethysmography devices are being researched \cite{charlton_2023_2023}, including sleep staging \cite{Kotzen_2022}, mental health assessment \cite{cakmak_classification_2021, Lyzwinski_2023}, identifying obstructive sleep apnea \cite{Behar_2014_Sleepap, Behar_2019feasibility}, and detecting peripheral arterial disease \cite{stansby_prospective_2022}. Each of these applications uses PPG signal analysis to derive physiological information from the PPG.

Photoplethysmography confers several advantages over other physiological monitoring technologies, which has resulted in its widespread adoption. Measurements can be obtained quickly without the need for a trained operator, and photoplethysmography sensors are non-invasive, unobtrusive, and low-cost. It is also a more compact and accessible monitoring modality than other measurements such as electrocardiogram (ECG) and blood pressure measurement. Furthermore, PPG measurements can be obtained without significantly disrupting daily activities, while ECG electrodes for instance may require careful placement and proper skin preparation to ensure accurate readings. However, a key disadvantage is that the PPG signal is highly susceptible to noise, such as in the cases of poor sensor contact or motion \cite{Li_2012}.



\subsection{Standardising PPG signal analysis}

A key step in the use of photoplethysmography for health and fitness monitoring is the development of PPG signal analysis algorithms. Such algorithms typically extract either inter-beat-intervals (\textit{e.g.}, for detection of atrial fibrillation) or features of PPG pulse wave shape (\textit{e.g.}, for estimation of blood pressure). However, unlike in other fields such as heart rate variability analysis, there are no standards for PPG signal analysis, and only limited open tools are available. Consequently, standardized and reproducible analysis of PPG signals is lacking.  Although there are some open-source PPG toolboxes, they lack validation and are often incomplete (see Table \ref{tab_PPG_tools}).

Despite the extensive research and applications in the field of PPG analysis, there is an urgent need to standardize approaches, terminologies, variables and definitions. Furthermore, there is no comprehensive toolbox available that covers all standard PPG biomarkers. It is important to acknowledge that certain variables may have different terminologies in the scientific literature, but our objective was to unify them to facilitate a broader understanding of PPG biomarkers. To fill this gap, we have developed a standardized nomenclature and toolbox. The assigned names for variables aim to provide insights into their origin, while the definitions ensure accurate interpretation and improved comprehensibility.

\subsection{Paper overview}
\begin{figure*}[!ht]
  \centering
  \includegraphics[width=\textwidth]{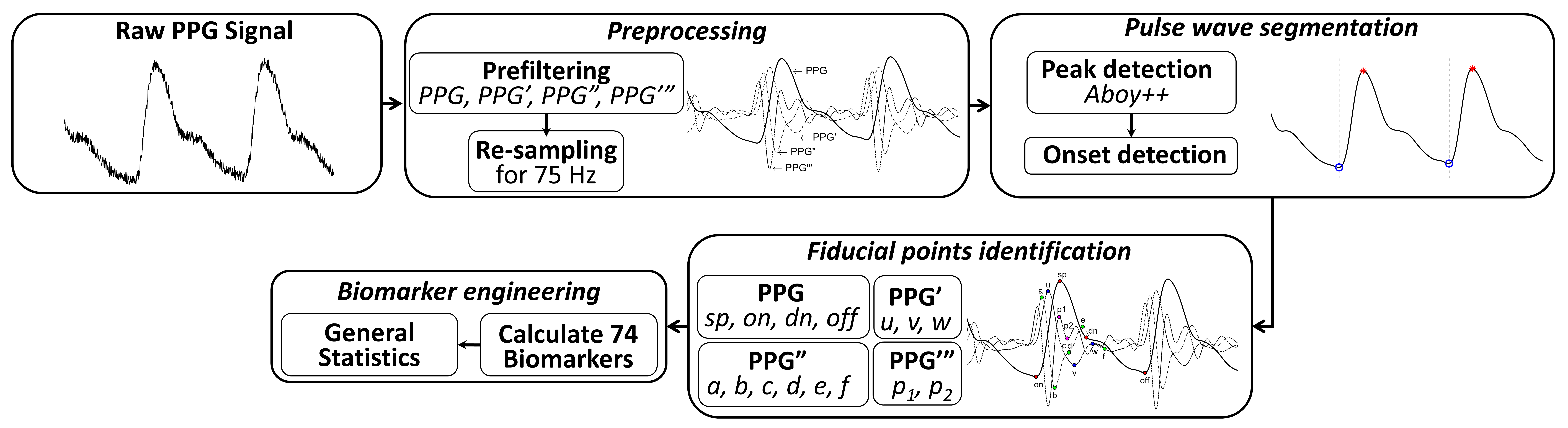}
  \caption{\scriptsize{Flowchart for continuous PPG time series analysis. The analysis comprises several key components, including: Preprocessing, Pulse wave segmentation, Fiducial points identification and Biomarker engineering.}}
  \label{fig:pipeline}
\end{figure*}

The primary aim of this research was to create a standardized toolbox ($pyPPG$) to analyze long-term finger PPG recordings in real-time. This paper presents standardized definitions for the state-of-the-art PPG fiducial points and biomarkers implemented within the $pyPPG$ toolbox. It provides an overview of the steps involved in raw data processing and biomarker engineering, as well as a validation of the fiducial point extraction process (see Figure \ref{fig:pipeline}). Additionally, the paper presents performance results and benchmarks them against other publicly available toolboxes.

\begin{table}[!hb]
\begin{center}
\fontsize{7pt}{7pt}
\selectfont
\caption{Comparison of open-source PPG signal processing libraries: pyPPG (this work), PPGFeat \cite{Abdullah_2023ppgfeat}, PulseAnalyse \cite{Charlton2019}, NeuroKit2 \cite{makowski2021_neurokit2}, RRest \cite{Data_2016}, PPGSynth \cite{Tang_2020ppgsynth}, PhysioNet Cardiovascular Signal Toolbox (PCST) \cite{Vest_2018}, HeartPy \cite{Gent_2019, Gent_2019analysing}, 
BioSPPy \cite{Carreiras_2015biosppy}}\label{tab_PPG_tools}
\addtolength{\tabcolsep}{-3.5pt}
\fontsize{7pt}{9pt}
\begin{tabular}{l|ccccccccc}
      & \multicolumn{1}{l}{\textbf{\hyperlink{https://github.com/aim-lab/GODA_pyPPG}{pyPPG}}}
      & \multicolumn{1}{l}{\textbf{\hyperlink{https://github.com/saadsur/PPGFeat}{PPGFeat}}}
      & \multicolumn{1}{l}{\textbf{\hyperlink{https://peterhcharlton.github.io/pulse-analyse/}{PulseAnalyse}}} & \multicolumn{1}{l}{\textbf{\hyperlink{https://github.com/neuropsychology/NeuroKit/tree/master/neurokit2/ppg}{NeuroKit2}}} & \multicolumn{1}{l}{\textbf{\hyperlink{https://github.com/peterhcharlton/RRest}{RRest}}} & \multicolumn{1}{l}{\textbf{\hyperlink{https://github.com/Elgendi/PPG-Synthesis/tree/master/code}{PPGSynth}}} & \multicolumn{1}{l}{\textbf{\hyperlink{https://physionet.org/content/pcst/1.0.0/}{PCST}}} & \multicolumn{1}{l}{\textbf{\hyperlink{https://github.com/paulvangentcom/heartrate_analysis_python}{HeartPy}}} & \multicolumn{1}{l}{\textbf{\hyperlink{https://github.com/PIA-Group/BioSPPy/}{BioSPPy}}}\\ \hline
\textit{Prefiltering}             & \Checkmark & \Checkmark & \Checkmark & \Checkmark   & \Checkmark   & \Checkmark   & \Checkmark & \Checkmark & \Checkmark\\
\textit{Peak detection}           & \Checkmark& \Checkmark & \Checkmark & \Checkmark   & \Checkmark   & \Checkmark   & \Checkmark & \Checkmark & \Checkmark\\
\textit{Onset detection}          & \Checkmark& \Checkmark & \Checkmark & \Checkmark   & \Checkmark    & –    & \Checkmark & – & \Checkmark\\
\textit{Other fiducial points}    & \Checkmark& \Checkmark & \Checkmark & –    & –    & –    & – & –  & –\\
\textit{Biomarker engineering}   & \Checkmark&  – & \Checkmark  & –    & –    & –    & – & –  & –\\
\textit{Signal quality}           & \Checkmark&  – & \Checkmark & –    & \Checkmark   & \Checkmark   & \Checkmark & \Checkmark & –\\
\textit{Quantitative validation}  & \Checkmark&  \Checkmark & –  & –    & –    & –    & \Checkmark & \Checkmark  & –                                              
\end{tabular}
\end{center}
\end{table}

\section{Materials and methods}\label{sec_method}

\subsection{Databases}\label{method_database}
Two databases were used to validate the $pyPPG$ toolbox. The Multi-Ethnic Study of Atherosclerosis (MESA) database \cite{Dean_2016, Zhang_2018} was used to validate our peak detector, and the PPG and Blood Pressure (PPG-BP) database \cite{Liang_2018} was used to validate the fiducial point detection algorithm. The MESA database consists of polysomnography (PSG) recordings from 2,056 adults aged 54–95 years, with subclinical cardiovascular disease, including 19,998 hours of PPG recordings \cite{Chen_2015, Rinkevivcius_2023}. Males accounted for 45.5\% of the subjects. The database was downloaded from the National Sleep Resource Center \cite{Zhang_2018}. The institutional review board from the Technion-IIT Rapport Faculty of Medicine was obtained under number 62-2019 in order to use the retrospective databases obtained from the open-access sleepdata.org resource for this research. The PSG recordings in MESA were acquired at home, including fingertip PPG signals measured at 256 Hz from the fingertip using Nonin 8000 series pulse oximeters (Nonin Medical Inc., Plymouth, US), alongside simultaneous ECG signals. The PPG-BP dataset contains 657 short (2 s) PPG recordings collected from 219 adult subjects aged 20–89 years with different health statuses (including healthy, hypertensive and diabetic subjects). Males accounted for 48\% of the subjects. The data include fingertip PPG signals measured at 1 kHz using a SEP9AF-2 PPG sensor (SMPLUS Company, Korea). Signals were acquired using a 12-bit ADC, and the hardware applied a 0.5‒12Hz band-pass filter.

\begin{table}[!hbt]
\setlength{\tabcolsep}{3pt}
\fontsize{7pt}{9pt}
\selectfont
\begin{center}
\begin{minipage}{350pt}
\caption{Fingertip PPG databases used for the quantitative validation experiments}\label{tab: DB}
\begin{tabular}{ccccccc}
\toprule
Database & Number of subjects & Length of recordings & Gender(M:F)    & Filtering & Sampling rate & Age\\\midrule
MESA    & 2056  & $\sim$10-hour & 1:1.2 & digital   & 256 Hz& 54-95\\
PPG-BP   & 219    & 2-second  & 1:1.08 & hardware    & 1 kHz & 20-89\\
\bottomrule
\end{tabular}
\end{minipage}
\end{center}
\end{table}

\subsection{Overview of the pyPPG toolbox}\label{method_pyPPG}

The \textit{pyPPG} toolbox is a standardized resource for real-time analysis of long-term finger PPG recordings. The toolbox consists of five main components, as summarized in Figure \ref{fig:pipeline}:
\begin{enumerate}
    \item \textbf{Loading a raw PPG signal:} The toolbox can accept various file formats such as $.mat$, $.csv$, $.txt$, or $.edf$. These files should contain raw PPG data along with the corresponding sampling rate.
    \item \textbf{Preprocessing:} The raw signal is filtered to remove unwanted noise and artifacts. Subsequently, the signal is resampled to a uniform rate of 75 Hz.
    \item \textbf{Pulse wave segmentation:} The toolbox employs a peak detector to identify the systolic peaks. Based on the peak locations, the toolbox also detects the pulse onsets and offsets, which indicate the start and end of the PPG pulse waves.
    \item \textbf{Fiducial points identification:} For each pulse wave, the toolbox detects a set of fiducial points.
    \item \textbf{Biomarker engineering:} Based on the fiducial points, a set of 74 PPG digital biomarkers are engineered.
\end{enumerate}

The $pyPPG$ toolbox also provides an optional PPG signal quality index based on the Matlab implementation of the work by Li \textit{et al.} \cite{Li_2012}. 


\subsection{Preprocessing}\label{method_prefilt}
The PPG signal filtering is one of the most essential parts of preprocessing. The human heart rate ranges between 30 and 200 beats per minute \cite{Paliakaite_2020}. Therefore in PPG signal analysis, it is common to apply bandpass filtering such: 0.5$-$8 Hz \cite{Abdullah_2023ppgfeat}, 0.5$-$10 Hz \cite{Finnegan_2023}, 0.5$-$15 Hz \cite{Mejia_2022}, 0.5$-$20 Hz \cite{Allen_2000, Liang_2018}, or 0.5$-$25 Hz \cite{Chowdhury2020}, to conserve the frequency content of the PPG pulse waves while filtering out lower frequency content (e.g. baseline wander due to respiration) and higher frequency content (e.g. muscle noise or power interference).

We selected a frequency range of 0.5$-$12 Hz. Whilst fiducial point detection can be simpler with lower low-pass cut-off frequencies such as 8 Hz, the drawback of using lower cut-off frequencies is that they significantly distort the pulse wave shape and reduce the accuracy with which the pulse onset and other fiducial points can be identified. Conversely, cut-off frequencies above 12 Hz can make it more complex to detect fiducial points due to the presence of extra waves in the PPG derivatives. Therefore, during the benchmarking process of other toolboxes for fiducial point detection (see Section \ref{results_Benchmark}), the 0.5$-$12 Hz frequency band was employed for filtering purposes. Although the 0.5$-$12 Hz band is recommended by default for PPG analysis, it is possible for the user to customize the passband filter in the \textit{pyPPG} toolbox.

To meet the above-mentioned requirements, the following zero-phase filters were implemented:
\begin{enumerate}
    \item \textbf{Bandpass filtering between 0.5$-$12 Hz:} A fourth-order Chebyshev Type II filter was used for the original signal. The 12 Hz low-pass cut-off was used to avoid time-shifting of fiducial points (particularly pulse onset, and dicrotic notch) and to eliminate unwanted high-frequency content from the PPG derivatives. The 0.5 Hz high-pass cut-off was used to minimize baseline wandering whilst retaining content at low heart rates.
    \item \textbf{20 ms moving average filtering (MAF):} In the case of very noisy signals, some high-frequency content can remain in the band-pass filter signal. For this purpose, a 20 ms standard flat (boxcar or top-hat) MAF with a 22.5 Hz cut-off frequency was applied after the band-pass filtering.
    \item \textbf{10 ms MAF for the PPG derivatives:} To eliminate the high-frequency content in the PPG derivatives, a 10 ms standard flat (boxcar or top-hat) MAF with 45 Hz cut-off frequency was applied. 
\end{enumerate}

It is common for the PPG signal to be sampled at over 100 Hz and up to 1 kHz, as, for example, in the PPG-BP dataset. However, an excessive sampling frequency may not be ideal for long-term data processing due to the computational load. The default behavior of the toolbox is to resample PPG signals at 75 Hz using the Python $resample$ function using the Fourier method.


\subsection{Pulse wave segmentation}\label{method_Segmentation}

The toolbox identifies individual pulse waves in a PPG signal by identifying systolic peaks ($sp$), and then identifying the pulse onset ($on$) and offset ($off$) on either side of each systolic peak which indicate the start and end of the pulse wave, respectively.

\begin{figure}[t]
    \centering
    \includegraphics[width=\columnwidth]{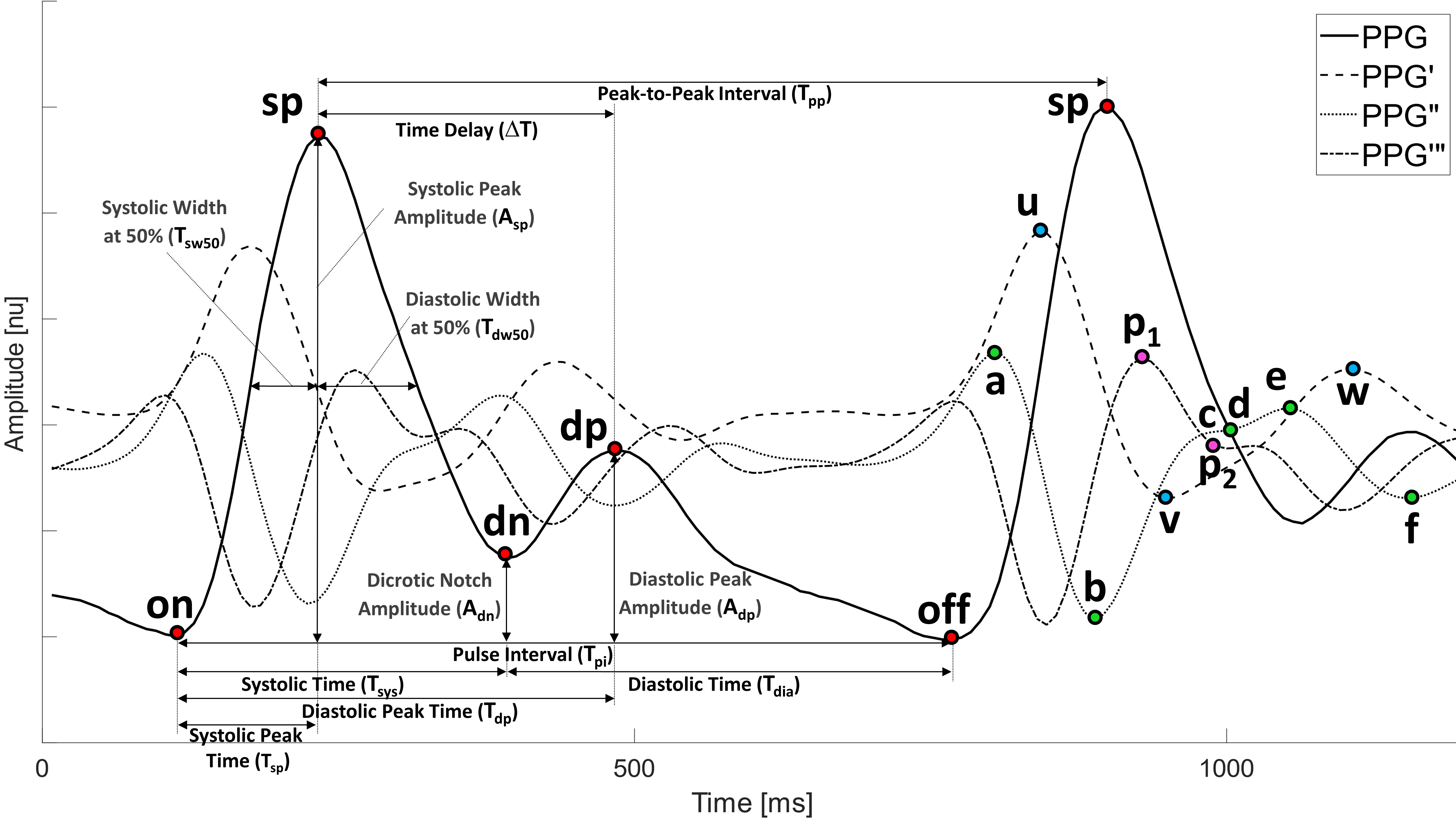}
    \caption{\scriptsize{The fiducial points of the PPG signal include the systolic peak ($sp$), the pulse onset and offset ($on$, $off$), the dicrotic notch ($dn$), and the diastolic peak ($dp$). The fiducial points of PPG derivatives are represented by $u$, $v$, $w$, $a$, $b$, $c$, $d$, $e$, $f$, $p_1$, $p_2$. The biomarkers are calculated based on this set of fiducial points.}}
    \label{fig:fiducial_points}
\end{figure}

\begin{table}[!hbt]
\fontsize{7pt}{9pt}
\selectfont
\begin{center}
\caption{Definition of PPG fiducial points}\label{tab:fiducial_points}

\addtolength{\tabcolsep}{-6pt}
\begin{tabular}{llllll}
\toprule
\multicolumn{5}{c}{\textbf{FIDUCIAL POINTS}} & \textbf{Ref}\\\midrule
\multicolumn{3}{c}{\textbf{PPG}}            & \multicolumn{3}{l}{} \\
1   & \textit{\textbf{on}} & \multicolumn{3}{l}{\textbf{\textit{Pulse onset}}. The beginning of the systolic upslope, typically, but not necessarily, a minimum point} & \\
2   & \textit{\textbf{sp}} & \multicolumn{3}{l}{\textbf{\textit{Systolic peak}}. The highest amplitude  between two consecutive pulse onsets} & \\ 
3   & \textit{\textbf{dn}} & \multicolumn{3}{l}{\begin{tabular}[c]{@{}l@{}}\textbf{\textit{Dicrotic notch}}. If a diastolic peak is present, then it is the local minimum preceding the diastolic peak. \\If there is no diastolic peak, then it is the inflection point between the systolic peak and f-point\end{tabular}}\\ 
4   & \textit{\textbf{dp}} & \multicolumn{3}{l}{\begin{tabular}[c]{@{}l@{}} \textbf{\textit{Diastolic peak}}. The first local maximum of the PPG pulse wave after the dicrotic notch and before the 0.8 pulse interval;\\ if no maxima then the first local maximum of PPG’ pulse wave after the e-point and before the 0.8 pulse interval.\end{tabular}}&\cite{Takazawa_1998}\\
5   & \textit{\textbf{off}}  & \multicolumn{3}{l}{\textbf{\textit{Pulse offset}}: The local minimum   preceding the next pulse wave's systolic upslope}\\\midrule

\multicolumn{2}{c}{\textbf{PPG'}}           & \multicolumn{3}{l}{}\\
5   & \textit{\textbf{u}}                   & \multicolumn{3}{l}{The highest amplitude   between the pulse onset and systolic peak on PPG'}  & \cite{Alty_2003}\\
6   & \textit{\textbf{v}}                   & \multicolumn{3}{l}{The lowest amplitude   between the u-point and diastolic peak on PPG'}&\cite{Suboh_2022}\\
7   & \textit{\textbf{w}}                   & \multicolumn{3}{l}{The first local maximum or inflection point after the dicrotic notch on PPG’}&\cite{Suboh_2022} \\\midrule

\multicolumn{3}{c}{\textbf{PPG"}}           & \multicolumn{3}{l}{}\\
8   & \textit{\textbf{a}}                   & \multicolumn{3}{l}{The highest amplitude   between pulse onset and systolic peak on PPG"}&\cite{Takazawa_1998}\\
9   & \textit{\textbf{b}}                   & \multicolumn{3}{l}{The first local   minimum after the a-point on PPG"}&\cite{Takazawa_1998}\\
10  & \textit{\textbf{c}}                   & \multicolumn{3}{l}{\begin{tabular}[c]{@{}l@{}}The local maximum with   the highest amplitude between the b-point and e-point, \\      or if no local maximum is present then the inflection point on PPG"\end{tabular}}&\cite{Takazawa_1998}\\
11  & \textit{\textbf{d}}                   & \multicolumn{3}{l}{\begin{tabular}[c]{@{}l@{}}The local minimum with   the lowest amplitude between the c-point and e-point, \\      or if no local minimum is present then the inflection point on PPG"\end{tabular}}&\cite{Takazawa_1998}\\
12  & \textit{\textbf{e}}                   & \multicolumn{3}{l}{The local maximum with   the highest amplitude after the b-point and before the diastolic peak on PPG"}&\cite{Takazawa_1998}\\
13  & \textit{\textbf{f}}                   & \multicolumn{3}{l}{The first local   minimum after the e-point on PPG"}&\cite{Takazawa_1998}\\\midrule

\multicolumn{3}{c}{\textbf{PPG'"}}          & \multicolumn{3}{l}{}\\
14  & \textit{\textbf{p$_1$}}                  & \multicolumn{3}{l}{The first local   maximum after the b-point on PPG'"}&\cite{Charlton_2018assessing}\\
15  & \textit{\textbf{p$_2$}}                  & \multicolumn{3}{l}{The last local minimum after the b-point and before the d-point on PPG'"} &\\                                               
\bottomrule
\end{tabular}
\end{center}
\end{table}

\subsubsection*{Systolic peak detection}\label{method_peak}
The $sp$ is the most important fiducial point of the PPG signal (see Figure \ref{fig:fiducial_points}). It is defined as the point with the highest amplitude between two consecutive pulse onsets (see Fig. \ref{fig:fiducial_points}). The \textit{pyPPG} toolbox uses an enhanced $sp$ detection algorithm to enable real-time analysis of long-term PPG measurements. The algorithm is an enhanced version of the Aboy beat detector \cite{Aboy2005}, which performed either best \cite{Kotzen_2022_MSc}, or amongst the best \cite{charlton_2022detecting} in recent benchmarking studies of PPG beat detectors. We focused on improving the beat detector's performance and reducing its computational complexity.

The original $Aboy$ algorithm utilizes an advanced filtering technique to accurately detect systolic peaks \cite{Aboy2005}. PPG recordings are segmented into 10-second windows and then filtered using three digital filters. The first filter helps to estimate the heart rate, while the second and third filters are used for peak detection. Two modifications were made to the \textit{Aboy} algorithm \cite{Aboy2005}. Firstly, to enhance the speed of the previous Matlab implementation \cite{charlton_2022detecting}, the finite impulse response (FIR) filter was replaced by a zero-phase fifth-order Chebyshev Type II infinite impulse response (IIR) filter, which applied the same cut-off frequencies as the original $Aboy$ peak detector. Secondly, the enhanced algorithm includes adaptive heart rate estimation to handle strong baseline wandering and rapid amplitude fluctuations \cite{Goda_2023Robust}. The resulting modified peak detector is denoted \textit{Aboy++}.

\subsubsection*{Pulse onset detection}\label{method_sonset}

The $on$ corresponds to the beginning of the pulse wave and the beginning of the systolic upslope (see Figure \ref{fig:fiducial_points} and Table \ref{tab:fiducial_points}). This systolic upslope is caused by increasing arterial pressure during systole \cite{Addison2016}. $on$ is typically a minimum point, but not necessarily. $pyPPG$ includes a novel $on$ detection algorithm. Previously, $on$ has been identified as the minimum value between two successive detected $sp$ \cite{Vadrevu_2019, Farooq_2010}, or identified using the slope sum function approach \cite{Nemati_2016, Deshmane_2009}. However, during long-term measurements there can be multiple local minima between successive $sp$, particularly in a noisy PPG signal. We define the $on$ as the initiation of the systolic upslope, which is usually a minimum point, although not always. We used a simple yet accurate approach to detect the $on$ as the first maximum preceding the $p_1$-point on the PPG'". $off$ is equivalent to the $on$ on the next pulse wave.

\subsection{Fiducial points identification}

\subsubsection*{Dicrotic notch detection}\label{method_DN}

The dicrotic notch ($dn$) plays an important role as a fiducial point in the analysis of PPG signals, holding immense potential for various applications such as heart disease detection \cite{Gu_2008} and arterial stiffness assessment \cite{Addison2016}. Its significance stems from its association with the duration of systole, which is known to be affected by heart disease. Additionally, the appearance of the diastolic wave following the $dn$ allows for the evaluation of arterial stiffness, with the hypothesis that the presence of the $dn$ is influenced by the arterial stiffness. However, it should be noted that the visibility of the $dn$ diminishes progressively with age, making it typically no longer discernible in elderly subjects \cite{Charlton_2022assessing}.

\begin{figure}[!ht]
    \centering
    \includegraphics[width=\columnwidth]{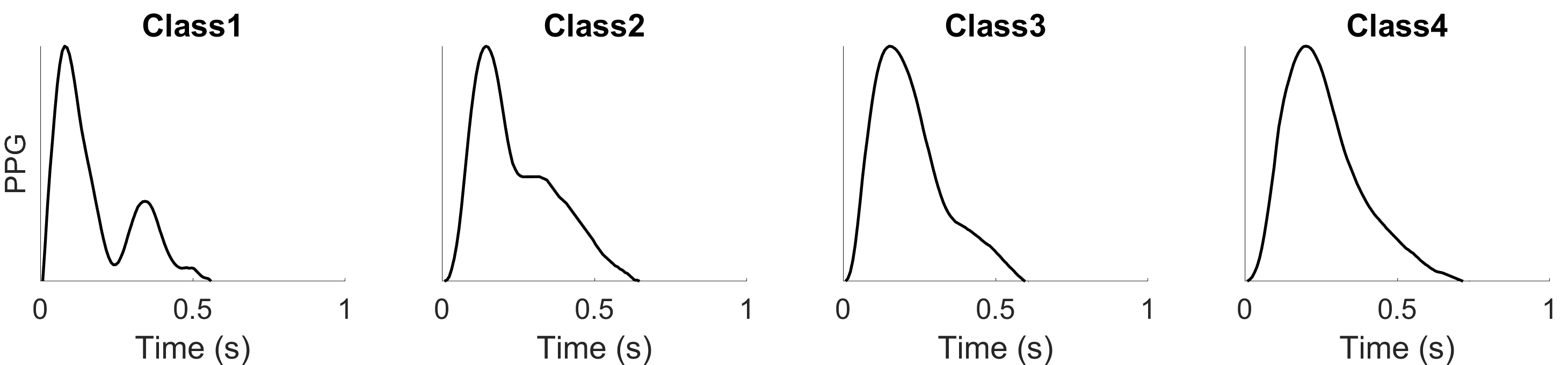}
    \caption{\scriptsize{
    The dicrotic notch ($dn$) is considered visible when there is a local minimum between the $sp$ and $dp$. However, the $dp$ is not always clearly visible. The $dn$ can be classified into four classes: Class 1, in which the $dn$ is with an incisura, Class 2, in which there is a horizontal line at the $dn$, Class 3, in which there is a change in gradient on the downslope, or and Class 4, in which there is no clear evidence of the $dn$. The figure was adapted from Charlton \textit{et al.} \cite{Charlton_2022assessing}.}}
    \label{fig:DN_classes}
\end{figure}

None of the existing definitions of $dn$ are entirely satisfactory. Typically, the $dn$ is easily recognizable when a distinct local minimum exists between the $sp$ and the $dp$ (see Figure \ref{fig:fiducial_points}). Yet, in many cases, the $dp$ is not clearly visible, rendering it difficult to accurately identify the $dn$. Dawber \textit{et al.} \cite{Dawber_1973} categorized different classes of $dn$, which are illustrated in Figure \ref{fig:DN_classes}. Another morphological approach for $dn$ identification involves locating it at the time of zero-crossing of the PPG'' between the \textit{d} and \textit{e} points \cite{Chakraborty_2021}. However, debate regarding the precise location for defining the $dn$ is ongoing. For instance, situations may arise where the local minimum of the $dn$ is visible, but the occurrence of the $d$ and $e$ points precedes the zero-crossing point, as depicted in Figure \ref{fig:fiducial_points}. 

%
\subsubsection*{Fiducial points of PPG derivatives}\label{method_fiduderiv}

Additional fiducial points were defined on the PPG derivatives (PPG', PPG'' and PPG'") as depicted in Figure \ref{fig:fiducial_points} \cite{Charlton_2022assessing, Suboh_2022}. The fiducial point detection algorithms in $pyPPG$ are based on standardized, morphological definitions (see Table \ref{tab:fiducial_points}). Consequently, these points do not necessarily correspond to points with consistent physiological interpretations. On the PPG' signal, the maximum point of the systolic slope is denoted as the \textit{u}-point, while the minimum point is referred to as the \textit{v}-point. The \textit{u}-point has been used to assess arterial stiffness \cite{Wowern_2015}. On the PPG'' signal, six further fiducial points are defined. Among these, four points ($a$, $b$, $c$, and $d$) are typically observed during the systolic phase (see Figure \ref{fig:fiducial_points}). As the diastolic phase begins, the $e$-point becomes visible, followed by the appearance of the $f$-point \cite{Suboh_2022}. Points $a$ to $e$ have been used to assess vascular ageing \cite{Takazawa_1998}, and the \textit{d}-point has been identified as a predictor of cardiovascular mortality. On the PPG'" signal, $p_1$ represents the early systolic component of the PPG pulse wave, while $p_2$ corresponds to the late systolic component \cite{Takazawa_1998}. $p_1$ and $p_2$ are used to calculate the augmentation index, which has been found to be elevated in atherosclerotic and diabetic subjects \cite{Bortolotto2000,Pilt2014}. 

\subsection{Biomarker engineering}\label{method_biomarker}

The \textit{pyPPG} toolbox includes a comprehensive collection of 74 standard PPG morphological biomarkers which are calculated from the timings and amplitudes of the fiducial points (see Tables \ref{tab: BM-PPGs}, \ref{tab: CBM-PPGs}, \ref{tab: BM-PPGd}, \ref{tab: CBM-PPGd}). The biomarkers were categorized into four groups: (1) PPG Signal - biomarkers that are based on the location of the fiducial points of the PPG signal; (2) Signal Ratios - biomarkers that are based on ratios of the fiducial points of the PPG signal; (3) PPG Derivatives - biomarkers that are based on the location of the fiducial points of the PPG derivatives; and (4) Derivatives Ratios - biomarkers that are based on ratios of the fiducial points of the PPG derivatives.

For a given window consisting of a set of beats, \textit{pyPPG} provides the following nine general statistics for each biomarker (see Supplementary Table \ref{tab_BM_stat_PPG}, \ref{tab_cBM_stat_PPG}, \ref{tab_BM_stat_PPGd}, \ref{tab_cBM_stat_PPGd}): signal duration; average (AVG); median (MED); standard deviation (STD); lower and upper quartiles (Q1, Q3); inter-quartile range (IQR); Skewness (SKW, indicating a lack of symmetry in the distribution; Kurtosis (KUR, indicating the pointedness of a peak in the distribution curve); and the average difference between the mean and each data value (MAD). 

\begin{table}[!ht]
\begin{center}
\caption{Biomarkers derived from the PPG signal categorized into intervals, amplitudes, and areas}\label{tab: BM-PPGs}
\addtolength{\tabcolsep}{2pt}
\fontsize{8pt}{10.5pt}
\selectfont
\begin{tabular}{lcllc}
\toprule

\multicolumn{4}{c}{\textbf{PPG SIGNAL}}  & \textbf{Ref} \\\midrule
1  & \multirow{10}{*}{Intervals}  & \textit{\textbf{T$_{pi}$}} & Pulse interval, the time between the pulse onset and pulse offset  & \cite{Chowdhury2020} \\
2  & & \textit{\textbf{T$_{pp}$}} & Peak-to-peak interval, the time between two consecutive systolic peaks  & \cite{Kurylyak_2013} \\
3  & & \textit{\textbf{T$_{sys}$}} & Systolic time, the time between the pulse onset and dicrotic notch  & \cite{Ahn_2017} \\
4  & & \textit{\textbf{T$_{dia}$}} & Diastolic time, the time between the dicrotic notch and pulse offset & \cite{Ahn_2017} \\
5  & & \textit{\textbf{T$_{sp}$}} & Systolic peak time, the time between the pulse onset and systolic peak  & \cite{Alty_2003} \\
6  & & \textit{\textbf{T$_{dp}$}} & Diastolic peak time, the time between the pulse onset and diastolic peak  & \cite{Chowienczyk_1999} \\
7  & & \textit{\textbf{$\Delta$T}} & Time delay, the time between the systolic peak and diastolic peak & \cite{Chowienczyk_1999}  \\

8  & & \textit{\textbf{T$_{swx}$}} & {\begin{tabular}[c]{@{}l@{}}Systolic width, the width at x\% of the systolic peak amplitude\\between the pulse onset and systolic peak\end{tabular}}& \cite{Kurylyak_2013} \\

9  & & \textit{\textbf{T$_{dwx}$}} & {\begin{tabular}[c]{@{}l@{}}Diastolic width, the width at x\% of the systolic peak amplitude\\between the systolic peak and pulse offset  \end{tabular}}& \cite{Kurylyak_2013} \\
10 & & \textit{\textbf{T$_{pwx}$}} & Pulse width, the sum of the systolic width and diastolic width at x\% & \cite{Kurylyak_2013} \\\midrule

11 & \multirow{4}{*}{Amplitudes}  & \textit{\textbf{A$_{sp}$}} & {\begin{tabular}[c]{@{}l@{}} Systolic peak amplitude, the difference in amplitude\\between the pulse onset and systolic peak\end{tabular}}  & \cite{Chua_2006} \\
12 & & \textit{\textbf{A$_{dn}$}} & {\begin{tabular}[c]{@{}l@{}}Dicrotic notch amplitude, the difference in amplitude\\between the pulse onset and dicrotic notch\end{tabular}}& \cite{Duan_2016} \\
13 & & \textit{\textbf{A$_{dp}$}} & {\begin{tabular}[c]{@{}l@{}} Diastolic peak amplitude, the difference in amplitude \\between the pulse onset and diastolic peak\end{tabular}}& \cite{Duan_2016} \\
14 & & \textit{\textbf{A$_{off}$}} & {\begin{tabular}[c]{@{}l@{}} Pulse onset amplitude, the difference in amplitude\\between the pulse onset and pulse offset\end{tabular}}& \\\midrule

15 & \multirow{3}{*}{Areas}  & \textit{\textbf{AUC$_{pi}$}}  & {\begin{tabular}[c]{@{}l@{}}Area under pulse interval curve, the area under the pulse wave\\between pulse onset and pulse offset  \end{tabular}}& \cite{Duan_2016} \\
16 & & \textit{\textbf{AUC$_{sys}$}} &{\begin{tabular}[c]{@{}l@{}} Area under systolic curve, the area under the pulse wave\\between the pulse onset and dicrotic notch  \end{tabular}}& \cite{Ahn_2017} \\
17 & & \textit{\textbf{AUC$_{dia}$}} &{\begin{tabular}[c]{@{}l@{}} Area under diastolic curve, the area under the pulse wave\\between the dicrotic notch and pulse offset \end{tabular}}& \cite{Ahn_2017} \\
\bottomrule
\end{tabular}
\end{center}
\end{table}

\begin{table}[!hb]
\begin{center}
\caption{Biomarkers derived from the signal ratios categorized into intervals, amplitudes, areas, and  combinations thereof}\label{tab: CBM-PPGs}

\addtolength{\tabcolsep}{-1.8pt}
\fontsize{8pt}{10pt}
\selectfont
\begin{tabular}{lcllc}
\toprule

\multicolumn{4}{c}{\textbf{SIGNAL RATIOS}} & \textbf{Ref} \\\midrule
1  & \multirow{6}{*}{Intervals} & \textit{\textbf{IPR}} & Instantaneous pulse rate, 60/T$_{pi}$  & \cite{Lueken_2017} \\
2  & & \textit{\textbf{T$_{sys}$/T$_{dia}$}}   & Ratio of the systolic time vs. the diastolic time & \cite{Ahn_2017} \\
3  & & \textit{\textbf{T$_{pwx}$/T$_{pi}$}} &{\begin{tabular}[c]{@{}l@{}} Ratio of the pulse width at x\% of the systolic peak amplitude\\vs. the pulse interval   \end{tabular}}& \cite{Chowdhury2020} \\
4  & & \textit{\textbf{T$_{pwx}$/T$_{ps}$}} &{\begin{tabular}[c]{@{}l@{}} Ratio of the pulse width at x\% of the systolic peak amplitude\\vs. the systolic peak time  \end{tabular}}& \cite{Chowdhury2020} \\
5  & & \textit{\textbf{T$_{dwx}$/T$_{swx}$}}   & Ratio of the diastolic width vs. the systolic width at x\% width  & \cite{Kurylyak_2013} \\
6  & & \textit{\textbf{T$_{sp}$/T$_{pi}$}}  & Ratio of the systolic peak time vs. the pulse interval   &  \\\midrule

7  & \multirow{2}{*}{Amplitudes}  & \textit{\textbf{A$_{sp}$/A$_{off}$}} & Ratio of the systolic peak amplitude vs. the pulse offset amplitude  &  \\
8  & & \textit{\textbf{A$_{dp}$/A$_{sp}$}}  & {\begin{tabular}[c]{@{}l@{}}Reflection index, ratio of the diastolic peak amplitude\\vs. the systolic peak amplitude   \end{tabular}}& \cite{Chowienczyk_1999} \\\midrule

9  & Areas  & \textit{\textbf{IPA}} & {\begin{tabular}[c]{@{}l@{}}Inflection point area, ratio of the area under diastolic curve\\vs. the area under systolic curve \end{tabular}}& \cite{Wang_2009} \\\midrule

10 & \multirow{3}{*}{Combined}  & \textit{\textbf{T$_{sp}$/A$_{sp}$}}  & Ratio of the systolic peak time vs. the systolic peak amplitude  & \cite{Liu_2021} \\
11 & & \textit{\textbf{A$_{sp}$/$\Delta$T}}  & Stiffness index, ratio of the systolic peak amplitude vs. the time delay & \cite{Millasseau_2002} \\
12 & & \textit{\textbf{A$_{sp}$/(T$_{pi}$-T$_{sp}$)}} &{\begin{tabular}[c]{@{}l@{}} Ratio of the systolic peak amplitude vs. the difference\\between the pulse interval and systolic peak time\end{tabular}}& \cite{Chowdhury2020} \\
\bottomrule
\end{tabular}
\end{center}
\end{table}

\begin{table}[ht]
\begin{center}
\caption{Biomarkers derived from the PPG derivatives}\label{tab: BM-PPGd}

\addtolength{\tabcolsep}{7pt}
\fontsize{8pt}{10pt}
\selectfont

\begin{tabular}{lcllc}
\toprule
\multicolumn{4}{c}{\textbf{PPG DERIVATIVES}}  & \textbf{Ref} \\\midrule
1  & \multirow{15}{*}{Intervals}  & \textit{\textbf{T$_{u}$}}  & u-point time, the time between the pulse onset and u-point  & ms \\
2  & & \textit{\textbf{T$_{u}$}}  & v-point time, the time between the pulse onset and v-point  & \cite{Suboh_2022} \\
3  & & \textit{\textbf{T$_{w}$}}  & w-point time, the time between the pulse onset and w-point  & \cite{Suboh_2022} \\
4  & & \textit{\textbf{T$_{a}$}}  & a-point time, the time between the pulse onset and a-point  & \cite{Suboh_2022} \\
5  & & \textit{\textbf{T$_{b}$}}  & b-point time, the time between the pulse onset and b-point  & \cite{Suboh_2022} \\
6  & & \textit{\textbf{T$_{c}$}}  & c-point time, the time between the pulse onset and c-point  & \cite{Suboh_2022} \\
7  & & \textit{\textbf{T$_{d}$}}  & d-point time, the time between the pulse onset and d-point  & \cite{Suboh_2022} \\
8  & & \textit{\textbf{T$_{e}$}}  & e-point time, the time between the pulse onset and e-point  & \cite{Suboh_2022} \\
9  & & \textit{\textbf{T$_{f}$}}  & f-point time, the time between the pulse onset and f-point  & \cite{Suboh_2022} \\
10 & & \textit{\textbf{T$_{b-c}$}} & b$-$c time, the time between the b-point and c-point  & \cite{Charlton_2018assessing} \\
11 & & \textit{\textbf{T$_{b-d}$}} & b$-$d time, the time between the b-point and d-point  & \cite{Charlton_2018assessing} \\
12 & & \textit{\textbf{T$_{p_1}$}} & p$_1$-point time, the time between the pulse onset and p$_1$-point  & \cite{Suboh_2022} \\
13 & & \textit{\textbf{T$_{p_2}$}} & p$_2$-point time, the time between the pulse onset and p$_2$-point  & \cite{Suboh_2022} \\
14 & & \textit{\textbf{T$_{p_1-dp}$}} & p$_1-$dia time, the time between the p$_1$-point and diastolic peak & \cite{Peltokangas_2017} \\
15 & & \textit{\textbf{T$_{p_2-dp}$}} & p$_2-$dia time, the time between the p$_2$-point and diastolic peak & \cite{Peltokangas_2017} \\
\bottomrule
\end{tabular}
\end{center}
\end{table}

\begin{table}[!hb]
\begin{center}
\caption{Biomarkers derived from the derivatives ratios categorized into intervals, amplitudes, areas, and combinations of these}\label{tab: CBM-PPGd}

\addtolength{\tabcolsep}{2pt}
\fontsize{8pt}{9pt}
\selectfont
\begin{tabular}{lcllc}
\toprule
\multicolumn{4}{c}{\textbf{DERIVATIVES RATIOS}}   & \textbf{Ref} \\\midrule
1  & \multirow{11}{*}{Intervals}  & \textit{\textbf{Tu/Tpi}} & Ratio of the u-point time vs. the pulse interval & \cite{Chowdhury2020} \\
2  & & \textit{\textbf{T$_{v}$/T$_{pi}$}} & Ratio of the v-point time vs. the pulse interval & \cite{Chowdhury2020} \\
3  & & \textit{\textbf{T$_{w}$/T$_{pi}$}} & Ratio of the w-point time vs. the pulse interval & \\
4  & & \textit{\textbf{T$_{a}$/T$_{pi}$}} & Ratio of the a-point time vs. the pulse interval & \cite{Chowdhury2020} \\
5  & & \textit{\textbf{T$_{b}$/T$_{pi}$}} & Ratio of the b-point time vs. the pulse interval & \cite{Chowdhury2020} \\
6  & & \textit{\textbf{T$_{c}$/T$_{pi}$}} & Ratio of the c-point time vs. the pulse interval &  \\
7  & & \textit{\textbf{T$_{d}$/T$_{pi}$}} & Ratio of the d-point time vs. the pulse interval &  \\
8  & & \textit{\textbf{T$_{e}$/T$_{pi}$}} & Ratio of the e-point time vs. the pulse interval &  \\
9  & & \textit{\textbf{T$_{f}$/T$_{pi}$}} & Ratio of the f-point time vs. the pulse interval &  \\
10 & & \textit{\textbf{(T$_{u}$-T$_{a}$)/T$_{pi}$}} &{\begin{tabular}[c]{@{}l@{}} Ratio of the difference between the u-point time\\and a-point time vs. the pulse interval \end{tabular}}& \cite{Chowdhury2020} \\
11 & & \textit{\textbf{(T$_{v}$-T$_{b}$)/T$_{pi}$}} &{\begin{tabular}[c]{@{}l@{}} Ratio of the difference between the v-point time\\and b-point time vs. the pulse interval \end{tabular}}& \cite{Chowdhury2020} \\\midrule

12 & \multirow{17}{*}{Amplitudes} & \textit{\textbf{A$_{u}$/A$_{sp}$}} & Ratio of the u-point amplitude vs. the systolic peak amplitude & \cite{Alty_2003} \\
13 & & \textit{\textbf{A$_{v}$/A$_{u}$}}  & Ratio of the v-point amplitude vs. the u-point amplitude &  \\
14 & & \textit{\textbf{A$_{w}$/A$_{u}$}}  & Ratio of the w-point amplitude vs. the u-point amplitude &  \\
15 & & \textit{\textbf{A$_{b}$/A$_{a}$}}  & Ratio of the b-point amplitude vs. the a-point amplitude & \cite{Takazawa_1998} \\
16 & & \textit{\textbf{A$_{c}$/A$_{a}$}}  & Ratio of the c-point amplitude vs. the a-point amplitude & \cite{Takazawa_1998} \\
17 & & \textit{\textbf{A$_{d}$/A$_{a}$}}  & Ratio of the d-point amplitude vs. the a-point amplitude & \cite{Takazawa_1998} \\
18 & & \textit{\textbf{A$_{e}$/A$_{a}$}}  & Ratio of the e-point amplitude vs. the a-point amplitude & \cite{Takazawa_1998} \\
19 & & \textit{\textbf{A$_{f}$/A$_{a}$}}  & Ratio of the f-point amplitude vs. the a-point amplitude &  \\
20 & & \textit{\textbf{A$_{p_2}$/A$_{p_1}$}}  & Ratio of the p$_2$-point amplitude vs. the p$_1$-point amplitude & \cite{Peltokangas_2017} \\
21 & & \textit{\textbf{(A$_{c}$-A$_{b}$)/A$_{a}$}}  &{\begin{tabular}[c]{@{}l@{}}  Ratio of the difference between the b-point amplitude\\and c-point amplitude vs. the a-point amplitude   \end{tabular}}& \cite{Ahn_2017} \\
22 & & \textit{\textbf{(A$_{d}$-A$_{b}$)/A\textbf{}}}  &{\begin{tabular}[c]{@{}l@{}} Ratio of the difference between the b-point amplitude\\and d-point amplitude vs. the a-point amplitude \end{tabular}}& \cite{Ahn_2017} \\
23 & & \textit{\textbf{AGI}} & Aging index, (A$_b$-A$_c$-A$_d$-A$_e$)/A$_{a}$ & \cite{Takazawa_1998} \\
24 & & \textit{\textbf{AGI$_{mod}$}} & Modified aging index, (A$_{b}$-A$_{c}$-A$_{d}$)/A$_{a}$  & \cite{Ushiroyama_2005} \\
25 & & \textit{\textbf{AGI$_{inf}$}} & Informal aging index, (A$_{b}$-A$_{e}$)/A$_{a}$  & \cite{Baek_2007} \\
26 & & \textit{\textbf{AI}}  & Augmentation index, (PPG(Tp2)-PPG(Tp1))/Asp & \cite{Takazawa_1998} \\
27 & & \textit{\textbf{RI$_{p_1}$}} & Reflection index of p$_1$, A$_{dp}$/(PPG(T$_{p_1}$)-PPG(T$_{pi}$(0)))  & \cite{Peltokangas_2017} \\
28 & & \textit{\textbf{RI$_{p_2}$}} & Reflection index of p$_2$, A$_{dp}$/(PPG(p$_{2}$)-PPG(T$_{pi}$(0))) & \cite{Peltokangas_2017} \\\midrule

29 & \multirow{2}{*}{Combined}  & \textit{\textbf{SC}}  & Spring constant, PPG"(T$_{sp}$)/((A$_{sp}$-A$_{u}$)/A$_{sp}$) & \cite{Wei2013} \\
30 & & \textit{\textbf{IPAD}} &{\begin{tabular}[c]{@{}l@{}} Inflection point area plus normalised d-point amplitude,\\AUC$_{dia}$/AUC$_{sys}$+A$_{d}$/A$_{a}$  \end{tabular}}& \cite{Ahn_2017}\\ 
\bottomrule
\end{tabular}
\end{center}
\end{table}

\subsection{Validation} \label{method_validation}

\subsubsection*{Systolic peak detection} 

The performance and computational complexity of the enhanced $sp$ detection algorithm, \textit{Aboy++}, were evaluated. Performance was assessed in comparison to reference ECG-derived beats using the $F_1$-score, which is a commonly used statistic for evaluating the performance of such algorithms. The $F_1$-score is particularly suitable for this purpose because it effectively combines multiple fractional measures by utilizing a harmonic mean between the sensitivity and positive predictive value. $F_1$-scores are reported as median (MED) and quartiles (Q1, Q3). The performance and computational complexity of \textit{Aboy++} were compared against the implementation of \textit{Aboy} provided by Charlton \textit{et al.} \cite{charlton_2022detecting}. Due to the high computational needs of \textit{Aboy}, the two algorithms were compared on a subset of MESA consisting of 100 recordings (1,173 hours) of PPG. \textit{Aboy++} was then assessed on the entire MESA database, with the exception of two recordings which did not have an ECG reference signal. Thus 2,054 PPG recordings were included, consisting of more than 19,000 hours of continuous PPG signals and over 91 million reference beats. The median recording length was 10 hours, with a 2.5-hour interquartile range (IQR). The ~10-hour-long recordings were divided into 10-minute segments. Segments were excluded if they did not contain a minimum of 300 ECG reference beats or if the extracted biomarkers could not be successfully evaluated. A key step in this assessment was to synchronise the timings of ECG-derived beats and PPG systolic peaks. This was achieved by forecasting the PPG $sp$ by extracting electrocardiogram (ECG) peaks from the PSG recordings as a reference signal similar to the work of Kotzen \textit{et al.}  \cite{Kotzen_2021}. The evaluation metric was based on the alignment of the ECG-R-wave and PPG $sp$. The methods for performance assessment are elaborated in more detail in our previous work \cite{Goda_2023Robust,Kotzen_2021}. 


\subsubsection*{Fiducial point detection}

The fiducial point detection algorithm was validated by comparison against the manual annotations of the PPG-BP \cite{Liang_2018} database. The data were manually annotated by two annotators (MAG and PHC) following the definitions in Table \ref{tab:fiducial_points}. An annotation tool was adapted from the open source \href{RRest}{https://github.com/peterhcharlton/RRest} \href{toolbox}{https://github.com/peterhcharlton/RRest} for this purpose \cite{charlton_2017extraction}. If an annotator could not confidently identify a fiducial point then they did not annotate it. After both annotators independently annotated the prefiltered signal, the time difference between the two annotators annotations was calculated. If the time discrepancy was \textgreater 10 ms, then the annotators discussed the case and either: (i) agreed on a location; or (ii) excluded the fiducial point (i.e. the annotators were not confident of its location). The final reference annotations were determined as the average of the annotations provided by the two annotators. In the PPG-BP database, each subject has three recordings. The first complete, high-quality pulse wave was selected for each subject. In total, more than 3,000 fiducial points from 219 patients were manually annotated by the two annotators. The PPG signals were filtered within the frequency band of 0.5-12 Hz during the manual annotation and benchmarking of the toolboxes (see Section \ref{results_Benchmark}). The Inter-rater reliability of annotations is presented in Table \ref{tab:annot_diff}.

To assess the performance of the fiducial point detection algorithm, $pyPPG$ was benchmarked against two publicly available PPG toolboxes capable of detecting fiducial points ($PulseAnalyse$ \cite{Charlton2019} and $PPGFeat$ \cite{Abdullah_2023ppgfeat}). Both benchmarked toolboxes were implemented in Matlab.  Performance was assessed using the mean absolute error (MAE) and the standard deviation of the absolute errors (STD) of the fiducial point detections in comparison to the reference . Bland-Altman plots \cite{Bland_1986} with the limits of agreement (1.96STD, indicating 95\% of errors) are also included.

\section{Results}\label{sec_results}
\subsection{Systolic peak detection}\label{results_peak}
The improved $sp$ detector, \textit{Aboy++}, outperformed the original \textit{Aboy} implementation when benchmarked on a subset of 100 randomly selected MESA recordings. The $F_1$-score of \textit{Aboy++} was $\sim$5\% greater than that of \textit{Aboy}. In addition, the computational time of \textit{Aboy++} was over 57 times faster than that of \textit{Aboy}. Specifically, the median peak detection time for 1-hour-long segments was 114.24 seconds for \textit{Aboy}, compared to 1.98 seconds for \textit{Aboy++}. When evaluated on the 2,054 recordings of the MESA dataset, \textit{Aboy++} obtained an $F_1$-score of 88.19\% (81.73$-$92.71).
 

\subsection{Fiducial points detection}\label{results_Benchmark}

\begin{figure}[ht]
    \centering
    \includegraphics[width=0.95\columnwidth]{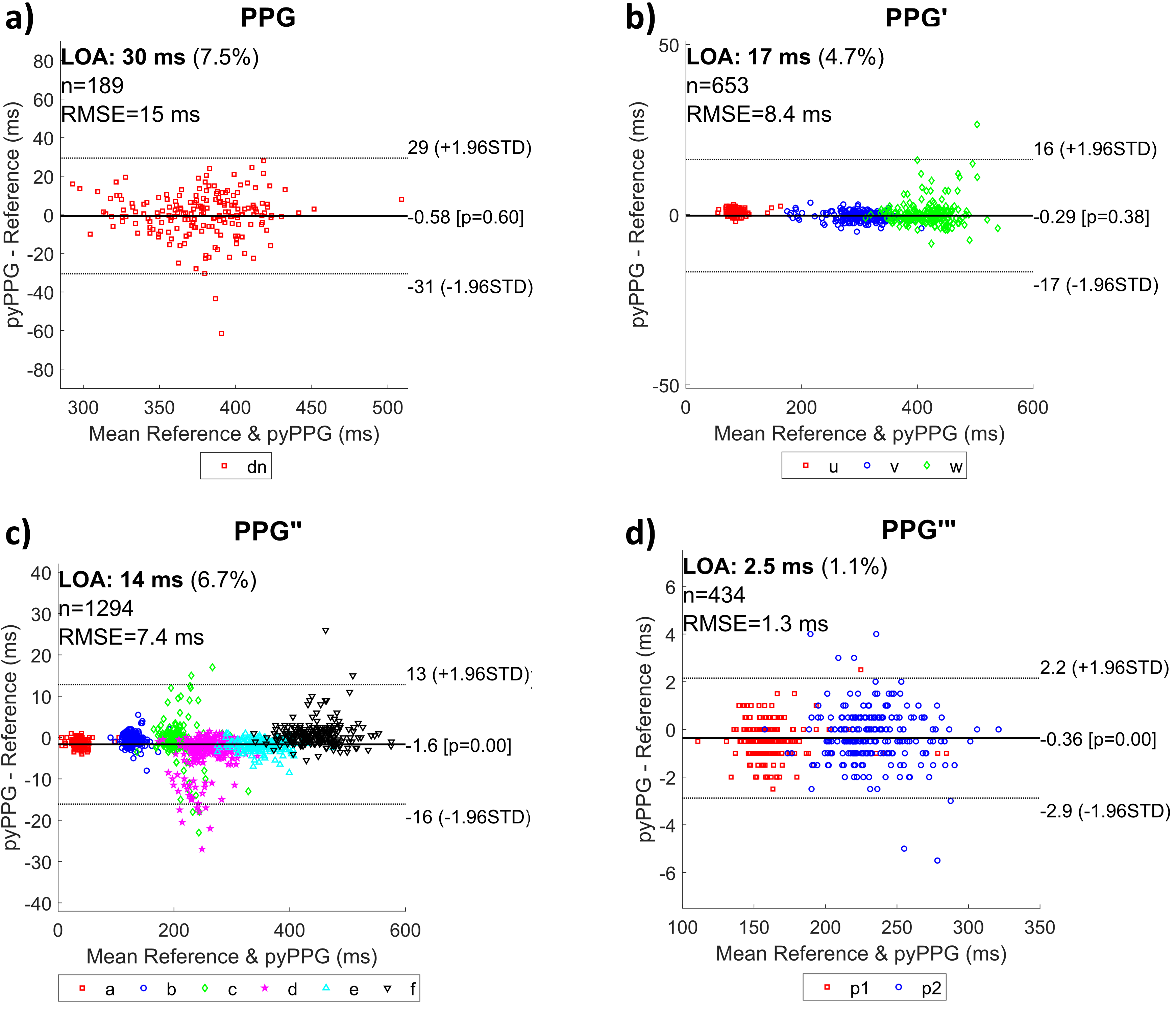}\label{figs: BlandAltman_MG-pyPPG}
    \caption{\scriptsize{Bland-Altman plot comparing the detected fiducial points and manual annotations at 99\% percentile of the data, with the reference pulse onset serving as the starting point. The panels (a) to (d) depict the results of the PPG signal and its derivatives.LOA: limits of agreement, n: number of fiducial points, RMSE: root mean squared error, STD: standard deviation of the differences.}}
    \label{fig:BlandAltman_MG-pyPPG}
\end{figure}

The results for the benchmarking of $pyPPG$ against other PPG toolboxes ($PulseAnalyse$ and $PPGFeat$) are presented in Table \ref{tab: Benchmarked toolboxes}. A total of 219 distinct pulse waves were employed for the benchmarking process. The MAEs were \textless 10 ms with $pyPPG$ for all fiducial points. $pyPPG$ outperformed the other toolboxes: The MAEs for $pyPPG$ were less than half of those for $PPGFeat$ for all fiducial points, and less than those for $PulseAnalyse$ for all except two fiducial points. In comparison to the other toolboxes, $pyPPG$ showed particular improvements in the detection of $dn$, $p_1$, and $p_2$. In addition, $pyPPG$ was able to detect fiducial points which were not detected by $PulseAnalyse$ ($v$ and $w$) or $PPGFeat$ ($p_1$, $p_2$). A standard Bland-Altman plot was generated to present the differences between the manual annotations and the $pyPPG$ fiducial points detection (see Figure \ref{figs: BlandAltman_MG-pyPPG}).

\begin{table}[t]
\fontsize{7pt}{9pt}
\selectfont

\begin{center}
\begin{minipage}{450pt}
\caption{Benchmark of PPG toolboxes for fiducial points detection. The mean (and standard deviation) of the absolute errors are reported for each fiducial in ms. }\label{tab: Benchmarked toolboxes}
\addtolength{\tabcolsep}{-3pt}
\begin{tabular}{cccccccccccccc}
\toprule
\textbf{Fiducial   Point} & \textbf{$on$} & \textbf{$dn$} & \textbf{$u$} & \textbf{$v$} & \textbf{$w$} & \textbf{$a$} & \textbf{$b$} & \textbf{$c$} & \textbf{$d$} & \textbf{$e$} & \textbf{$f$} & \textbf{$p_1$} & \textbf{$p_2$}\\
\midrule

\textbf{\textit{\textbf{pyPPG (this work)}}}& \textbf{8(13)} & \textbf{9(15)} & \textbf{1(1)} & \textbf{1(1)} & \textbf{3(14)} & \textbf{1(1)} & \textbf{1(1)} & \textbf{3(6)} & \textbf{5(9)} & \textbf{3(9)} & \textbf{3(11)} & \textbf{1(1)} & \textbf{1(2)}\\
\textbf{\textit{\textbf{PulseAnalyse \cite{Charlton2019}}}}& \textbf{8(25)} & 26(23) & 2(1) & - & - & \textbf{1(1)} & 2(1) & 9(26) & 9(27) & 3(18) & 5(21) & 39(36) & 30(42)\\
\textbf{\textit{\textbf{PPGFeat \cite{Abdullah_2023ppgfeat}}}} & 50(139) & 40(56) &9(16) &13(38)& 48(99) & 16(32) & 10(21) & 19(36) & 23(45) &28(59) & 39(77) & - & -  \\

\bottomrule
\end{tabular}
\end{minipage}

\fontsize{7pt}{7pt}
\selectfont

\end{center}
\end{table}

\subsection{pyPPG and PhysioZoo PPG}\label{discuss_PhysioZoo}
The resulting systolic peak detection and fiducial points detection algorithms are package into an open source Python library denoted $pyPPG$. In addition, a user-friendly interface is also implemented in the \href{PhysioZoo}{https://physiozoo.com/}  \href{Software}{https://physiozoo.com/}. In order to ensure that $pyPPG$ could process a large dataset without technical issues, we run it over the full MESA database and reported standard statistics for all biomarkers (see Supplementary Table \ref{tab_BM_stat_PPG}, \ref{tab_cBM_stat_PPG}, \ref{tab_BM_stat_PPGd}, \ref{tab_cBM_stat_PPGd}). (The library and software will be placed online following publication.)

\section{Discussion}\label{sec_discuss}
This work is expected to contribute significantly to the scientific field of computerized cardiology, leading to a better understanding of the PPG signal. Firstly, it standardized the definition of PPG fiducial points and biomarkers. The second major contribution consists in the implementation and quantitative validation of a fiducial point detector. The peak detection algorithm in the $pyPPG$ toolbox was validated on 19,000 hours of continuous PPG data, encompassing more than 91 million reference beats. It performed with an 88.18\% $F_1$-score while processing a 1-hour segment in 1.98 seconds. When evaluated on 3,000 manually annotated fiducial points $pyPPG$ had a low MAE and consistently outperformed two other open toolboxes.

The third key contribution pertains to the implementation in $pyPPG$ of 74 physiological PPG. The toolbox is made open-source, making it the only comprehensive and validated Python library that is publicly accessible. A user-friendly interface is also implemented in the \href{PhysioZoo}{https://physiozoo.com/}  \href{Software}{https://physiozoo.com/} software. This interface enables data visualization, exploration and quantitative analysis of a PPG recording. By offering this novel solution, researchers and clinicians are empowered with a valuable resource for comprehensive and reproducible PPG analysis. Finally, the manual annotations of the 219 recordings, including more than 3,000 fiducial points are made open access to ensure reproducibility of the results and to enable further investigations and advancements in the field of PPG analysis. 

This research has limitations. The performance of the peak detection algorithms were only performed on sleep data from atherosclerosis patients (see Supplementary Table \ref{tab_BM_stat_PPG}, \ref{tab_cBM_stat_PPG}, \ref{tab_BM_stat_PPGd}, \ref{tab_cBM_stat_PPGd}). Therefore the evaluation of Aboy++ on additional databases would be very beneficial. Another limitation of the work is the focus on the analysis of PPG measured using standard clinical oximeter. Adapting the toolbox to incorporate PPG sources like earlobe PPG or smartwatches will also be of interest, particularly given the widespread use of the later. The program has another limitation related to the controversial nature of morphological and physiological characteristics of fiducial points (see Supplementary Figure \ref{figs:good_bad_fidu}). Hence, creating a standardized toolbox presented a significant challenge.

In conclusion, this work provides a standards and advanced toolbox for the analysis of PPG. Studying the PPG time series variability using \textit{pyPPG} can enhance our understanding of the manifestations and etiology of diseases. This toolbox can also be used for biomarker engineering in training data-driven models.




\section{Code Availability}\label{sec_availability}
The source code used and the annotations of the fiducial points in this research are available at physiozoo.com.
\section{Acknowledgments}\label{sec_acknowledgments}
MAG and JAB acknowledge the Estate of Zofia (Sophie) Fridman and funding from the Israel Innovation Authority. PHC acknowledges funding from the British Heart Foundation (grant FS/20/20/34626).

The Multi-Ethnic Study of Atherosclerosis (MESA) Sleep Ancillary study was funded by NIH-NHLBI Association of Sleep Disorders with Cardiovascular Health Across Ethnic Groups (RO1 HL098433). MESA is supported by NHLBI funded contracts HHSN268201500003I, N01-HC-95159, N01-HC-95160, N01-HC-95161, N01-HC-95162, N01-HC-95163, N01-HC-95164, N01-HC-95165, N01-HC-95166, N01-HC-95167, N01-HC-95168 and N01-HC-95169 from the National Heart, Lung, and Blood Institute, and by cooperative agreements UL1-TR-000040, UL1-TR-001079, and UL1-TR-001420 funded by NCATS. The National Sleep Research Resource was supported by the National Heart, Lung, and Blood Institute (R24 HL114473, 75N92019R002).


\bibliographystyle{plain}
\bibliography{refs}


\section{Supplementary Material}\label{sec_supplement}
\setcounter{table}{0}
\renewcommand{\thetable}{S\arabic{table}} 

\setcounter{figure}{0}
\renewcommand{\thefigure}{S\arabic{figure}} 

\begin{figure*}[!ht]
  \centering
  \includegraphics[width=\textwidth]{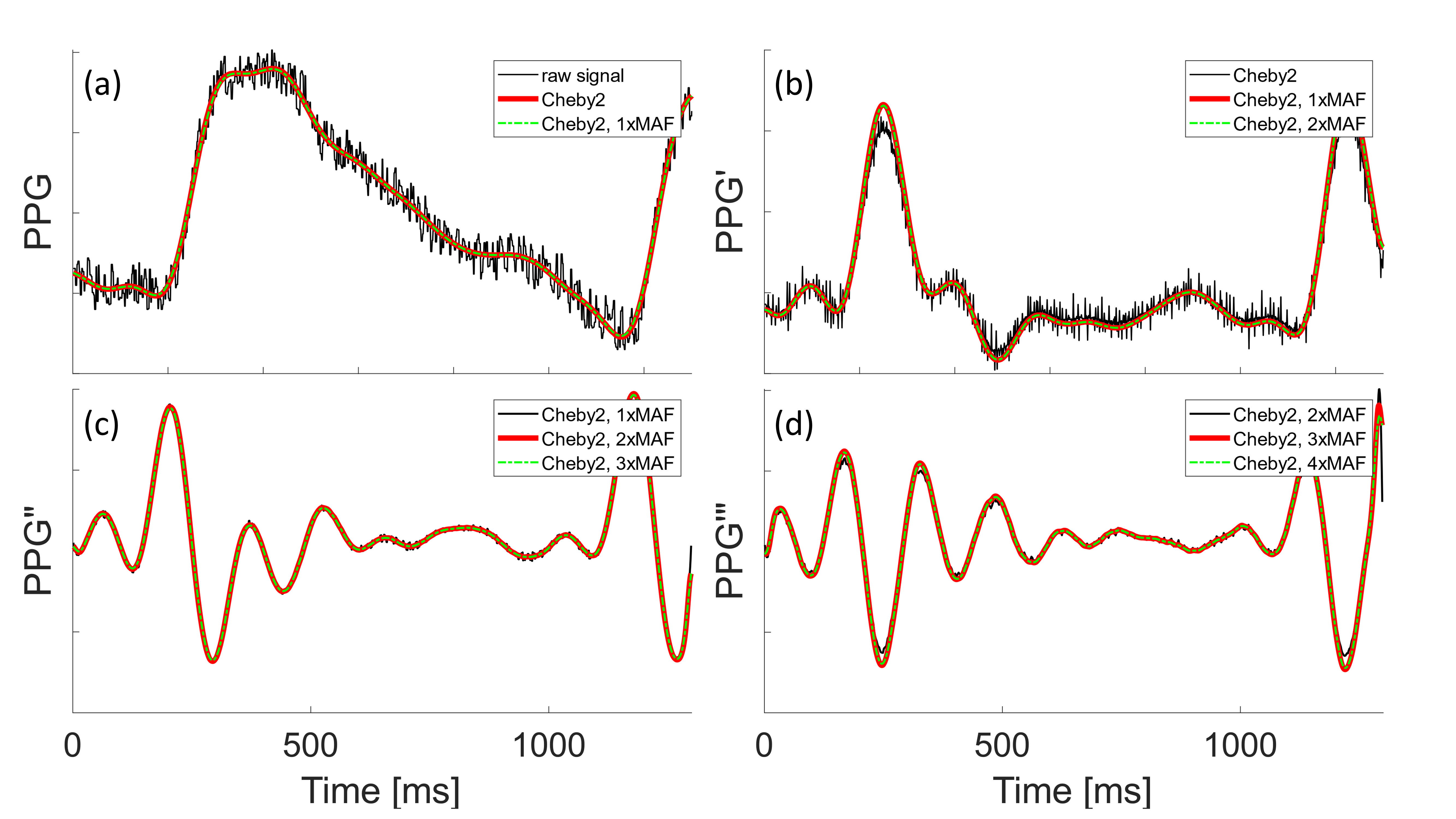}
  \caption{\scriptsize{Prefiltering of the PPG, PPG’, PPG’’, and PPG’’’ signals. In panel (a), the raw PPG signal is represented by the black curve, while the red curve represents the application of a fourth-order Chebyshev Type II filter (Cheby2). In panels (a) to (d), the green dashed curve represents the filtered version of the red curve obtained through moving average filtering (MAF). Panels (b) to (d) display the PPG', PPG'', and PPG'" signals, respectively. The black curve depicts the derivatives of the red curve from the preceding panel. Likewise, the red curve corresponds to the derivative of the green curve shown in the previous panel.}}
  \label{fig: filtering}
\end{figure*}

\begin{table}[!ht]
\fontsize{7pt}{9pt}
\selectfont

\begin{center}
\begin{minipage}{420pt}
\caption{Inter-annotator differences for fiducial points}\label{tab:annot_diff}

\addtolength{\tabcolsep}{-1pt}
\begin{tabular}{ccccccccccccccc}
\toprule

\textbf{Fiducial   Point} & \textbf{$sp$} & \textbf{$on$} & \textbf{$dn$} & \textbf{$u$} & \textbf{$v$} & \textbf{$w$} & \textbf{$a$} & \textbf{$b$} & \textbf{$c$} & \textbf{$d$} & \textbf{$e$} & \textbf{$f$} & \textbf{$p_1$} & \textbf{$p_2$} \\\midrule
\textbf{MAE (STD) ms}& 3(4)  & 2(3)  & 3(5)  & 1(1)  & 2(2)  & 2(3)  & 2(2)  & 3(2)  & 3(3)  & 2(3)  & 3(2)  & 2(3)  & 1(1) & 1(2)  \\

\bottomrule
\end{tabular}
\begin{tablenotes}
   \item The mean and standard deviation of the absolute errors (MAE and STD respectively) are reported for each fiducial in ms.
\end{tablenotes}

\end{minipage}
\end{center}
\end{table}

\begin{figure}[hbt]
    \centering
    \includegraphics[width=\columnwidth]{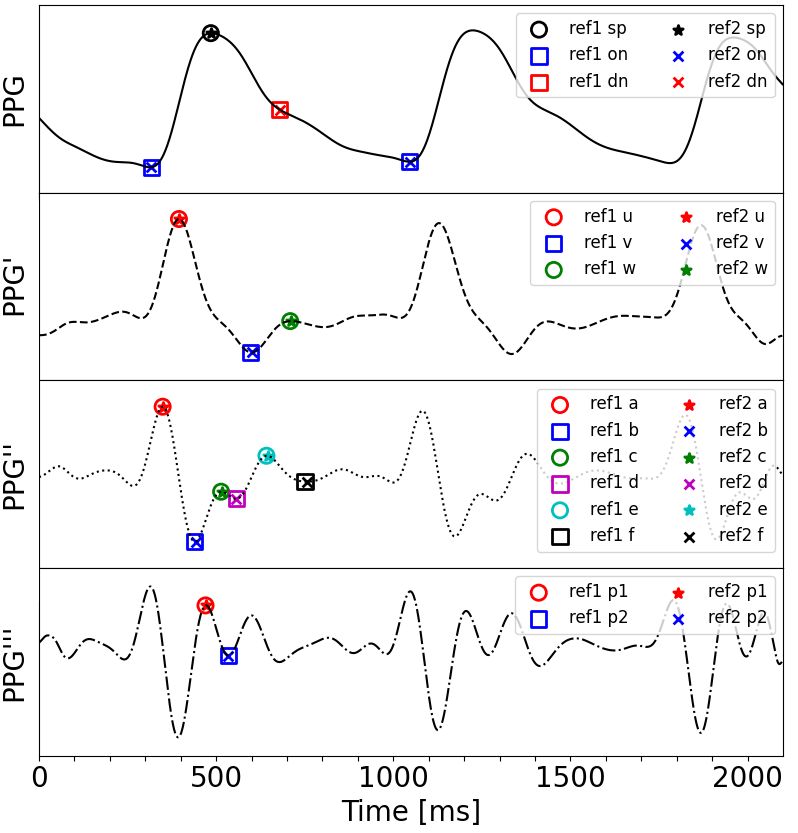}
    \caption{\scriptsize{Comparison of fiducial point annotations. The first manual reference annotation is represented by empty circles and squares, while the  crosses and stars indicate the mark of the second annotator. The black lines illustrate the PPG signal and its derivatives. More than 3000 fiducial points were manually annotated in 219 different PPG waves by two annotators on the PPG-BP dataset to validate the implemented fiducial points detector}}
    \label{fig:annot_fiducial_points}
\end{figure}

\begin{figure}[hb]
    \centering
    \includegraphics[width=\columnwidth]{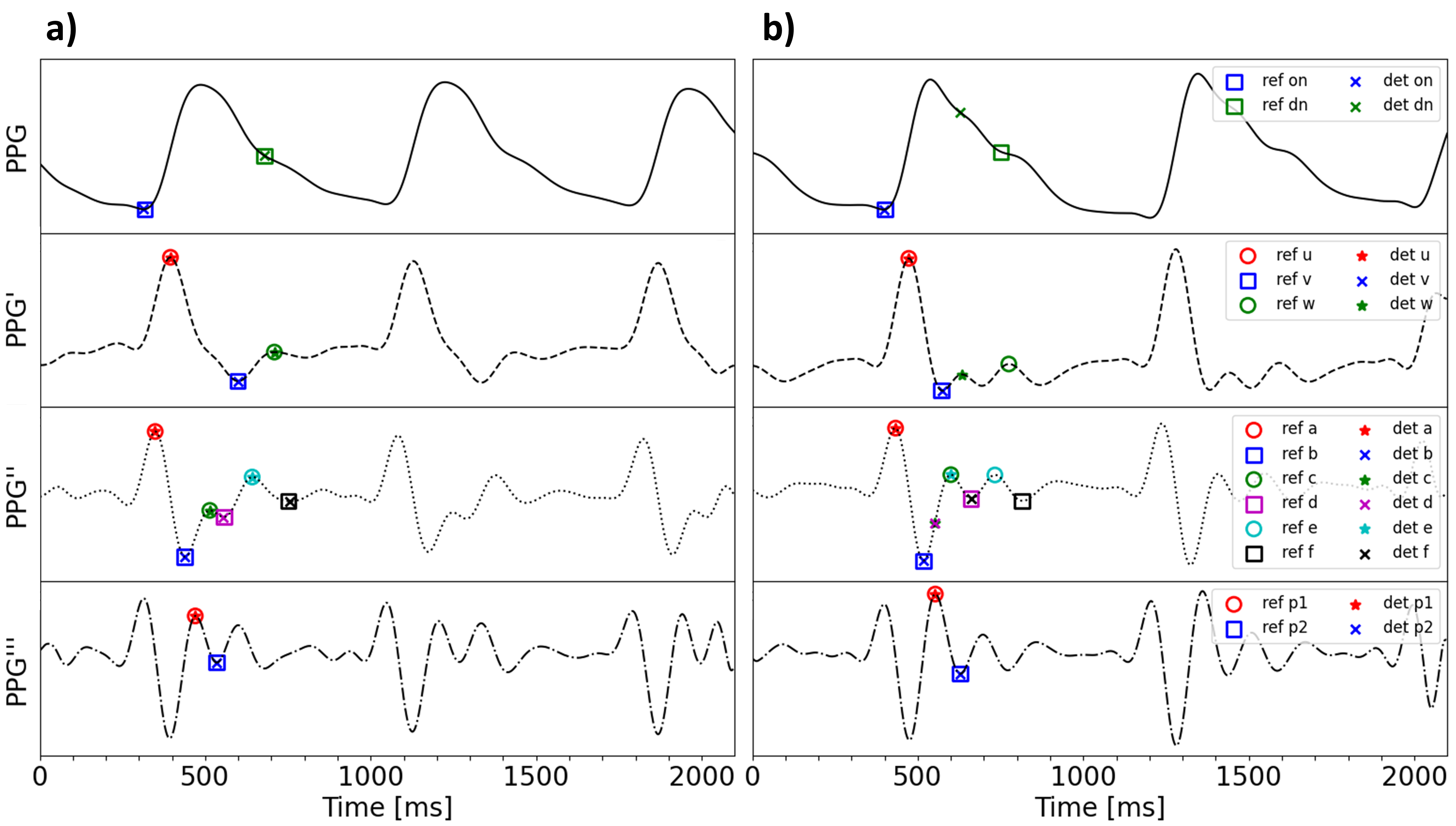}
    \caption{\scriptsize{A comparison between high and low-quality PPG signals. In panel (a), a high-quality signal is illustrated, where all fiducial points are clearly detectable. However, in panel (b), the fiducial point detections are inaccurate due to the low quality of the signal. The reference fiducial points are represented by empty circles and squares, while the detected points are indicated by crosses and stars.}}
    \label{figs: good_bad_fidu}
\end{figure}

\begin{table}[!hbt]
\begin{center}
\caption{Summary statistics for biomarkers of PPG signal from the MESA database for 2,054 PPG recordings}\label{tab_BM_stat_PPG}

\addtolength{\tabcolsep}{2pt}

\begin{tabular} {lllllllllll}
\toprule
\multicolumn{11}{c}{\textbf{PPG SIGNAL}}\\
\textbf{BIOMARKER}       & \textbf{AVG} & \textbf{MED} & \textbf{STD} & \textbf{Q1} & \textbf{Q3} & \textbf{IQR} & \textbf{SKW} & \textbf{KUR} & \textbf{MAD} & \textbf{Unit} \\ \midrule
\textit{\textbf{T$_{pi}$}}     	&	0.85 & 0.83 & 0.15  & 0.79  & 0.88  & 0.10 & 3.19  & 29.44  & 0.09  & s  \\
\textit{\textbf{T$_{pp}$}}     	&	0.85 & 0.83 & 0.14  & 0.79  & 0.88  & 0.09 & 2.71  & 24.63  & 0.08  & s  \\
\textit{\textbf{T$_{sys}$}}    	&	0.38 & 0.36 & 0.07  & 0.34  & 0.39  & 0.05 & 4.08  & 38.56  & 0.04  & s  \\
\textit{\textbf{T$_{dia}$}}    	&	0.47 & 0.47 & 0.13  & 0.41  & 0.51  & 0.10 & 1.89  & 20.18  & 0.08  & s  \\
\textit{\textbf{T$_{sp}$}}     	&	0.21 & 0.20 & 0.05  & 0.18  & 0.22  & 0.04 & 4.50  & 46.99  & 0.03  & s  \\
\textit{\textbf{T$_{dp}$}}     	&	0.38 & 0.36 & 0.07  & 0.34  & 0.39  & 0.05 & 4.08  & 38.56  & 0.04  & s  \\
\textit{\textbf{$\Delta$T}}       	&	0.17 & 0.16 & 0.05  & 0.14  & 0.19  & 0.05 & 2.12  & 14.02  & 0.03  & s  \\
\textit{\textbf{T$_{sw10}$}}   	&	0.16 & 0.16 & 0.04  & 0.14  & 0.18  & 0.04 & 2.96  & 27.86  & 0.03  & s  \\
\textit{\textbf{T$_{sw25}$}}   	&	0.14 & 0.14 & 0.04  & 0.12  & 0.16  & 0.03 & 2.68  & 26.06  & 0.02  & s  \\
\textit{\textbf{T$_{sw33}$}}   	&	0.13 & 0.13 & 0.04  & 0.11  & 0.15  & 0.03 & 2.57  & 25.51  & 0.02  & s  \\
\textit{\textbf{T$_{sw50}$}}   	&	0.11 & 0.11 & 0.03  & 0.10  & 0.13  & 0.03 & 2.32  & 23.75  & 0.02  & s  \\
\textit{\textbf{T$_{sw66}$}}   	&	0.10 & 0.09 & 0.03  & 0.08  & 0.11  & 0.03 & 2.12  & 22.20  & 0.02  & s  \\
\textit{\textbf{T$_{sw75}$}}   	&	0.08 & 0.08 & 0.03  & 0.07  & 0.10  & 0.03 & 2.11  & 22.22  & 0.02  & s  \\
\textit{\textbf{T$_{sw90}$}}   	&	0.06 & 0.05 & 0.02  & 0.04  & 0.06  & 0.02 & 2.45  & 23.71  & 0.01  & s  \\
\textit{\textbf{T$_{dw10}$}}   	&	0.50 & 0.50 & 0.13  & 0.44  & 0.55  & 0.11 & 1.39  & 16.16  & 0.08  & s  \\
\textit{\textbf{T$_{dw25}$}}   	&	0.38 & 0.37 & 0.12  & 0.33  & 0.42  & 0.09 & 2.04  & 18.59  & 0.08  & s  \\
\textit{\textbf{T$_{dw33}$}}   	&	0.33 & 0.32 & 0.12  & 0.27  & 0.36  & 0.09 & 2.30  & 19.30  & 0.07  & s  \\
\textit{\textbf{T$_{dw50}$}}   	&	0.23 & 0.21 & 0.10  & 0.18  & 0.25  & 0.07 & 3.21  & 24.36  & 0.06  & s  \\
\textit{\textbf{T$_{dw66}$}}   	&	0.16 & 0.14 & 0.09  & 0.12  & 0.17  & 0.05 & 4.51  & 38.24  & 0.05  & s  \\
\textit{\textbf{T$_{dw75}$}}   	&	0.13 & 0.11 & 0.08  & 0.09  & 0.13  & 0.04 & 5.18  & 47.43  & 0.04  & s  \\
\textit{\textbf{T$_{dw90}$}}   	&	0.07 & 0.06 & 0.07  & 0.05  & 0.07  & 0.02 & 6.14  & 63.89  & 0.03  & s  \\
\textit{\textbf{T$_{pw10}$}}   	&	0.67 & 0.66 & 0.14  & 0.61  & 0.72  & 0.11 & 1.60  & 18.13  & 0.09  & s  \\
\textit{\textbf{T$_{pw25}$}}   	&	0.52 & 0.51 & 0.13  & 0.47  & 0.56  & 0.10 & 2.05  & 19.56  & 0.08  & s  \\
\textit{\textbf{T$_{pw33}$}}   	&	0.46 & 0.45 & 0.12  & 0.40  & 0.50  & 0.10 & 2.22  & 19.76  & 0.08  & s  \\
\textit{\textbf{T$_{pw50}$}}   	&	0.35 & 0.33 & 0.11  & 0.29  & 0.38  & 0.08 & 2.86  & 22.45  & 0.07  & s  \\
\textit{\textbf{T$_{pw66}$}}   	&	0.25 & 0.24 & 0.10  & 0.21  & 0.27  & 0.06 & 4.00  & 34.20  & 0.05  & s  \\
\textit{\textbf{T$_{pw75}$}}   	&	0.21 & 0.19 & 0.09  & 0.17  & 0.22  & 0.04 & 4.58  & 42.41  & 0.05  & s  \\
\textit{\textbf{T$_{pw90}$}}   	&	0.13 & 0.12 & 0.08  & 0.10  & 0.14  & 0.04 & 5.18  & 53.15  & 0.04  & s  \\
\textit{\textbf{A$_{sp}$}}     	&	0.20 & 0.18 & 0.12  & 0.14  & 0.22  & 0.07 & 3.50  & 34.29  & 0.06  & nu \\
\textit{\textbf{A$_{dn}$}}     	&	0.10 & 0.10 & 0.10  & 0.07  & 0.13  & 0.06 & 1.01  & 34.45  & 0.05  & nu \\
\textit{\textbf{A$_{dp}$}}     	&	0.09 & 0.08 & 0.09  & 0.06  & 0.11  & 0.05 & 2.31  & 34.98  & 0.05  & nu \\
\textit{\textbf{A$_{off}$}}    	&	0.00 & 0.00 & 0.07  & -0.01 & 0.01  & 0.02 & 1.03  & 40.12  & 0.03  & nu \\
\textit{\textbf{AUC$_{pi}$}}   	&	9.50 & 9.71 & 97.74 & 8.64  & 10.48 & 1.83 & -0.27 & 162.91 & 13.21 & nu \\
\textit{\textbf{AUC$_{sys}$}}  	&	4.37 & 4.33 & 23.77 & 3.76  & 4.97  & 1.20 & -0.24 & 148.02 & 3.49  & nu \\
\textit{\textbf{AUC$_{dia}$}}  	&	5.13 & 5.21 & 74.42 & 4.06  & 6.18  & 2.12 & -0.18 & 158.93 & 10.21 & nu \\

\bottomrule
\end{tabular}
\begin{tablenotes}
   \item  Average (AVG); median (MED); standard deviation (STD); lower and upper quartiles (Q1, Q3); inter-quartile range (IQR); Skewness (SKW, indicating a lack of symmetry in the distribution; Kurtosis (KUR, indicating the pointedness of a peak in the distribution curve); and the average difference between the mean and each data value (MAD).
\end{tablenotes}

\end{center}
\end{table}

\begin{table}[!hbt]

\begin{center}
\caption{Summary statistics for biomarkers of signal ratios from the MESA database for 2,054 PPG recordings}\label{tab_cBM_stat_PPG}

\addtolength{\tabcolsep}{0pt}

\begin{tabular} {lllllllllll}
\toprule
\multicolumn{11}{c}{\textbf{SIGNAL RATIOS}}\\
\textbf{BIOMARKER}              & \textbf{AVG} & \textbf{MED} & \textbf{STD} & \textbf{Q1} & \textbf{Q3} & \textbf{IQR} & \textbf{SKW} & \textbf{KUR} & \textbf{MAD} & \textbf{Unit} \\ \midrule
\textit{\textbf{IPR}}      	&	73.45   & 73.33   & 9.87    & 69.33   & 77.50   & 8.16   & 0.41    & 12.57    & 6.53   & \% \\
\textit{\textbf{T$_{sys}$/T$_{dia}$}}      	&	91.33   & 81.22   & 39.32   & 71.41   & 98.41   & 27.00  & 393.22  & 2923.76  & 24.38  & \% \\
\textit{\textbf{T$_{pw25}$/T$_{pi}$}}      	&	62.00   & 62.75   & 10.52   & 57.08   & 68.00   & 10.91  & -85.54  & 610.45   & 7.62   & \% \\
\textit{\textbf{T$_{pw50}$/T$_{pi}$}}      	&	40.87   & 39.97   & 9.85    & 35.34   & 45.27   & 9.93   & 78.04   & 627.29   & 7.09   & \% \\
\textit{\textbf{T$_{pw75}$/T$_{pi}$}}      	&	24.74   & 23.54   & 7.83    & 21.02   & 26.51   & 5.49   & 247.90  & 1915.65  & 4.88   & \% \\
\textit{\textbf{T$_{pw25}$/T$_{sp}$}}      	&	266.17  & 260.62  & 76.87   & 226.93  & 294.03  & 67.10  & 174.35  & 1698.59  & 50.18  & \% \\
\textit{\textbf{T$_{pw50}$/T$_{sp}$}}      	&	174.02  & 165.78  & 61.20   & 144.75  & 189.08  & 44.33  & 289.07  & 2596.04  & 36.91  & \% \\
\textit{\textbf{T$_{pw75}$/T$_{sp}$}}      	&	105.68  & 98.58   & 47.91   & 86.73   & 111.63  & 24.90  & 435.03  & 4669.79  & 24.71  & \% \\
\textit{\textbf{T$_{dw10}$/T$_{sw10}$}}    	&	336.94  & 323.76  & 132.82  & 271.42  & 380.80  & 109.38 & 250.74  & 2289.27  & 82.14  & \% \\
\textit{\textbf{T$_{dw25}$/T$_{sw25}$}}    	&	298.33  & 277.70  & 148.79  & 229.72  & 333.07  & 103.35 & 355.27  & 3273.42  & 84.17  & \% \\
\textit{\textbf{T$_{dw33}$/T$_{sw33}$}}    	&	277.89  & 252.94  & 155.82  & 207.23  & 308.78  & 101.55 & 400.58  & 3716.32  & 85.66  & \% \\
\textit{\textbf{T$_{dw50}$/T$_{sw50}$}}    	&	230.45  & 195.56  & 176.10  & 159.28  & 246.41  & 87.13  & 519.26  & 5062.43  & 86.84  & \% \\
\textit{\textbf{T$_{dw66}$/T$_{sw66}$}}    	&	197.33  & 154.47  & 212.89  & 123.35  & 199.78  & 76.43  & 634.82  & 6779.45  & 91.84  & \% \\
\textit{\textbf{T$_{dw75}$/T$_{sw75}$}}    	&	187.13  & 137.75  & 244.73  & 108.57  & 182.58  & 74.01  & 685.40  & 7569.23  & 99.48  & \% \\
\textit{\textbf{T$_{dw90}$/T$_{sw90}$}}    	&	185.76  & 114.54  & 373.62  & 90.54   & 154.42  & 63.88  & 800.79  & 9758.14  & 132.09 & \% \\
\textit{\textbf{T$_{sp}$/T$_{pi}$}} 	&	24.80   & 24.11   & 5.44    & 21.71   & 26.99   & 5.28   & 195.60  & 1503.32  & 3.80   & \% \\
\textit{\textbf{A$_{sp}$/A$_{off}$}}       	&	-251.50 & -225.29 & 5093.80 & -265.99 & -190.83 & 75.16  & -79.27  & 15171.47 & 673.97 & \% \\
\textit{\textbf{A$_{dp}$/A$_{sp}$}} 	&	50.65   & 47.32   & 306.54  & 37.37   & 56.42   & 19.05  & -132.42 & 8856.26  & 54.04  & \% \\
\textit{\textbf{IPA}}   	&	1.35    & 1.24    & 19.00   & 0.95    & 1.50    & 0.55   & 0.21    & 146.18   & 2.53   & nu \\
\textit{\textbf{T$_{sp}$/A$_{sp}$}} 	&	2.59    & 1.89    & 41.34   & 1.41    & 2.63    & 1.22   & 5.04    & 118.11   & 6.17   & nu \\
\textit{\textbf{A$_{sp}$/$\Delta$T}}    	&	1.32    & 1.13    & 1.06    & 0.87    & 1.48    & 0.61   & 4.26    & 43.19    & 0.54   & nu \\
\textit{\textbf{A$_{sp}$/(T$_{pi}$-T$_{sp}$)}} 	&	0.32    & 0.28    & 0.21    & 0.23    & 0.36    & 0.13   & 3.78    & 36.18    & 0.11   & nu\\

\bottomrule
\end{tabular}
\begin{tablenotes}
   \item  Average (AVG); median (MED); standard deviation (STD); lower and upper quartiles (Q1, Q3); inter-quartile range (IQR); Skewness (SKW, indicating a lack of symmetry in the distribution; Kurtosis (KUR, indicating the pointedness of a peak in the distribution curve); and the average difference between the mean and each data value (MAD).
\end{tablenotes}

\end{center}
\end{table}

\begin{table}[!hbt]

\begin{center}
\caption{Summary statistics for biomarkers of PPG derivative from the MESA database for 2,054 PPG recordings}\label{tab_BM_stat_PPGd}

\addtolength{\tabcolsep}{2pt}

\begin{tabular} {lllllllllll}
\toprule
\multicolumn{11}{c}{\textbf{PPG DERIVATIVES}}\\
\textbf{BIOMARKER}              & \textbf{AVG} & \textbf{MED} & \textbf{STD} & \textbf{Q1} & \textbf{Q3} & \textbf{IQR} & \textbf{SKW} & \textbf{KUR} & \textbf{MAD} & \textbf{Unit} \\ \midrule
\textit{\textbf{T$_{u}$}}     	&	0.11 & 0.08 & 0.11 & 0.06 & 0.11 & 0.05 & 6.21  & 62.13 & 0.06 & s \\
\textit{\textbf{T$_{v}$}}     	&	0.38 & 0.33 & 0.15 & 0.29 & 0.42 & 0.13 & 3.24  & 20.96 & 0.11 & s \\
\textit{\textbf{T$_{w}$}}     	&	0.42 & 0.37 & 0.16 & 0.32 & 0.47 & 0.14 & 2.80  & 15.51 & 0.11 & s \\
\textit{\textbf{T$_{a}$}}     	&	0.06 & 0.04 & 0.08 & 0.03 & 0.07 & 0.04 & 6.05  & 60.85 & 0.04 & s \\
\textit{\textbf{T$_{b}$}}     	&	0.11 & 0.09 & 0.08 & 0.08 & 0.12 & 0.04 & 5.33  & 50.57 & 0.05 & s \\
\textit{\textbf{T$_{c}$}}     	&	0.15 & 0.13 & 0.09 & 0.11 & 0.17 & 0.06 & 4.04  & 32.39 & 0.05 & s \\
\textit{\textbf{T$_{d}$}}     	&	0.22 & 0.21 & 0.11 & 0.16 & 0.27 & 0.11 & 2.02  & 11.56 & 0.07 & s \\
\textit{\textbf{T$_{e}$}}     	&	0.35 & 0.34 & 0.11 & 0.30 & 0.39 & 0.09 & 1.38  & 13.26 & 0.07 & s \\
\textit{\textbf{T$_{f}$}}     	&	0.38 & 0.37 & 0.11 & 0.33 & 0.43 & 0.10 & 1.33  & 11.52 & 0.08 & s \\
\textit{\textbf{T$_{b-c}$}}   	&	0.04 & 0.03 & 0.02 & 0.02 & 0.04 & 0.02 & 0.87  & 1.69  & 0.01 & s \\
\textit{\textbf{T$_{b-d}$}}   	&	0.11 & 0.10 & 0.07 & 0.06 & 0.15 & 0.09 & 1.28  & 5.24  & 0.05 & s \\
\textit{\textbf{T$_{p1}$}}    	&	0.13 & 0.11 & 0.08 & 0.09 & 0.14 & 0.05 & 4.92  & 44.11 & 0.05 & s \\
\textit{\textbf{T$_{p2}$}}    	&	0.21 & 0.19 & 0.10 & 0.14 & 0.25 & 0.11 & 2.31  & 13.10 & 0.07 & s \\
\textit{\textbf{T$_{p1-dp}$}} 	&	0.25 & 0.25 & 0.08 & 0.22 & 0.28 & 0.06 & -0.41 & 19.47 & 0.05 & s \\
\textit{\textbf{T$_{p2-dp}$}} 	&	0.17 & 0.17 & 0.10 & 0.11 & 0.23 & 0.11 & -0.57 & 7.90  & 0.07 & s \\
\bottomrule
\end{tabular}
\begin{tablenotes}
   \item  Average (AVG); median (MED); standard deviation (STD); lower and upper quartiles (Q1, Q3); inter-quartile range (IQR); Skewness (SKW, indicating a lack of symmetry in the distribution; Kurtosis (KUR, indicating the pointedness of a peak in the distribution curve); and the average difference between the mean and each data value (MAD).
\end{tablenotes}
\end{center}
\end{table}

\begin{table}[!hbt]

\begin{center}
\caption{Summary statistics for biomarkers of PPG derivative from the MESA database for 2,054 PPG recordings}\label{tab_cBM_stat_PPGd}

\addtolength{\tabcolsep}{0pt}

\begin{tabular} {lllllllllll}
\toprule
\multicolumn{11}{c}{\textbf{DERIVATIVES RATIOS}}\\
\textbf{BIOMARKER}              & \textbf{AVG} & \textbf{MED} & \textbf{STD} & \textbf{Q1} & \textbf{Q3} & \textbf{IQR} & \textbf{SKW} & \textbf{KUR} & \textbf{MAD} & \textbf{Unit} \\ \midrule
\textit{\textbf{T$_{u}$/T$_{pi}$}}      	&	11.18   & 8.44    & 11.57   & 7.32    & 9.95   & 2.63   & 498.54  & 2700.88  & 5.23   & \% \\
\textit{\textbf{T$_{v}$/T$_{pi}$}}      	&	49.82   & 41.88   & 19.96   & 34.71   & 63.47  & 28.76  & 82.65   & -12.92   & 16.85  & \% \\
\textit{\textbf{T$_{w}$/T$_{pi}$}}      	&	54.02   & 45.93   & 20.01   & 38.84   & 67.60  & 28.76  & 80.94   & -12.58   & 16.85  & \% \\
\textit{\textbf{T$_{a}$/T$_{pi}$}}      	&	5.80    & 4.07    & 7.96    & 2.59    & 5.52   & 2.94   & 547.31  & 3988.16  & 3.88   & \% \\
\textit{\textbf{T$_{b}$/T$_{pi}$}}      	&	11.42   & 10.28   & 8.04    & 8.44    & 11.72  & 3.28   & 539.86  & 4105.76  & 3.88   & \% \\
\textit{\textbf{T$_{c}$/T$_{pi}$}}      	&	15.07   & 14.05   & 8.46    & 11.35   & 16.66  & 5.31   & 469.69  & 3478.37  & 4.59   & \% \\
\textit{\textbf{T$_{d}$/T$_{pi}$}}      	&	25.59   & 23.16   & 12.71   & 16.50   & 32.05  & 15.55  & 134.11  & 393.65   & 9.83   & \% \\
\textit{\textbf{T$_{e}$/T$_{pi}$}}      	&	38.89   & 40.33   & 13.68   & 31.87   & 47.13  & 15.26  & -3.42   & 169.26   & 10.24  & \% \\
\textit{\textbf{T$_{f}$/T$_{pi}$}}      	&	43.02   & 44.63   & 13.63   & 35.81   & 51.29  & 15.48  & -2.06   & 162.95   & 10.28  & \% \\
\textit{\textbf{(T$_{u}$-T$_{a}$)/T$_{pi}$}} 	&	5.33    & 3.53    & 8.88    & 2.96    & 6.07   & 3.12   & 550.25  & 4378.08  & 3.95   & \% \\
\textit{\textbf{(T$_{v}$-T$_{b}$)/T$_{pi}$}} 	&	38.33   & 30.90   & 19.59   & 24.38   & 52.20  & 27.82  & 69.31   & -26.89   & 16.40  & \% \\
\textit{\textbf{A$_{u}$/A$_{sp}$}}	&	11.44   & 11.29   & 10.85   & 9.85    & 12.91  & 3.06   & -342.27 & 19223.85 & 3.01   & \% \\
\textit{\textbf{A$_{v}$/A$_{u}$}} 	&	-56.53  & -50.35  & 31.53   & -65.01  & -40.28 & 24.74  & -581.35 & 6066.96  & 17.48  & \% \\
\textit{\textbf{A$_{w}$/A$_{u}$}} 	&	8.39    & 6.60    & 21.72   & -5.48   & 21.43  & 26.91  & -23.67  & 268.64   & 16.59  & \% \\
\textit{\textbf{A$_{b}$/A$_{a}$}} 	&	-67.51  & -71.60  & 62.78   & -96.23  & -41.67 & 54.56  & -366.50 & 5557.12  & 39.35  & \% \\
\textit{\textbf{A$_{c}$/A$_{a}$}} 	&	8.73    & 6.21    & 42.85   & -14.34  & 30.65  & 44.99  & 151.81  & 1769.14  & 30.31  & \% \\
\textit{\textbf{A$_{d}$/A$_{a}$}} 	&	-75.06  & -67.10  & 68.71   & -94.36  & -48.42 & 45.95  & -515.47 & 7369.81  & 36.15  & \% \\
\textit{\textbf{A$_{e}$/A$_{a}$}} 	&	76.99   & 66.61   & 88.58   & 46.56   & 90.78  & 44.22  & 916.64  & 14471.11 & 32.28  & \% \\
\textit{\textbf{A$_{f}$/A$_{a}$}} 	&	-56.11  & -48.07  & 68.51   & -70.78  & -25.96 & 44.82  & -576.39 & 6746.56  & 33.40  & \% \\
\textit{\textbf{A$_{p2}$/A$_{p1}$}}     	&	29.51   & 127.29  & 1310.75 & 24.64   & 204.19 & 179.55 & -845.24 & 12641.48 & 370.00 & \% \\
\textit{\textbf{(A$_{c}$-A$_{b}$)/A$_{a}$}}	&	76.24   & 69.03   & 75.93   & 30.69   & 107.35 & 76.65  & 559.28  & 7767.55  & 46.26  & \% \\
\textit{\textbf{(A$_{d}$-A$_{b}$)/A$_{a}$}}	&	-7.55   & 6.35    & 69.52   & -33.05  & 34.80  & 67.84  & -152.71 & 442.67   & 50.05  & \% \\
\textit{\textbf{AGI}} 	&	-78.17  & -69.51  & 112.94  & -113.36 & -28.24 & 85.11  & -706.72 & 11560.28 & 57.14  & \% \\
\textit{\textbf{AGI$_{mod}$}}      	&	-1.18   & -9.10   & 66.72   & -44.11  & 42.13  & 86.23  & 45.44   & 152.19   & 52.43  & \% \\
\textit{\textbf{AGI$_{inf}$}}      	&	-144.50 & -130.56 & 132.13  & -172.72 & -99.70 & 73.01  & -831.10 & 13837.36 & 54.63  & \% \\
\textit{\textbf{AI}}  	&	42.33   & 38.33   & 84.63   & 17.14   & 57.40  & 40.26  & 924.79  & 13287.84 & 36.73  & \% \\
\textit{\textbf{RI$_{p1}$}} 	&	132.42  & 111.38  & 2657.66 & 83.44   & 154.65 & 71.21  & 325.91  & 17081.63 & 504.55 & \% \\
\textit{\textbf{RI$_{p2}$}} 	&	132.42  & 111.38  & 2657.66 & 83.44   & 154.65 & 71.21  & 325.91  & 17081.63 & 504.55 & \% \\
\textit{\textbf{SC}}  	&	0.00    & 0.00    & 0.00    & 0.00    & 0.00   & 0.00   & -16.01  & 271.47   & 0.00   & nu \\
\textit{\textbf{IPAD}} 	&	1.93    & 0.43    & 29.07   & -0.01   & 0.79   & 0.80   & 10.28   & 235.07   & 3.64   & nu\\
\bottomrule
\end{tabular}
\begin{tablenotes}
   \item  Average (AVG); median (MED); standard deviation (STD); lower and upper quartiles (Q1, Q3); inter-quartile range (IQR); Skewness (SKW, indicating a lack of symmetry in the distribution; Kurtosis (KUR, indicating the pointedness of a peak in the distribution curve); and the average difference between the mean and each data value (MAD).
\end{tablenotes}
\end{center}
\end{table}

\end{document}